\documentclass[aps,pre,epsf,superscriptaddress,amsmath,amssymb,amsfonts,showpacs,superscriptaddress,floatfix,preprintnumbers,twocolumn,10pt]{revtex4-1}
\usepackage{microtype}
\usepackage[export]{adjustbox}
\DisableLigatures[f]{encoding = *, family = *}
\usepackage{graphicx}
\usepackage{epsfig}
\usepackage{dcolumn}
\usepackage{bm}
\usepackage{braket}
\usepackage{amsmath}
\usepackage{mathtools}
\usepackage{color}
\newcommand{\abs}[1]{\left| #1 \right|}

\begin{document}

\title{Pattern formation of correlated impurities subjected\\ to an impurity-medium interaction pulse}

\author{G. Bougas}
\affiliation{Center for Optical Quantum Technologies, Department of Physics, University of Hamburg, 
Luruper Chaussee 149, 22761 Hamburg Germany}
\author{S. I. Mistakidis}
\affiliation{Center for Optical Quantum Technologies, Department of Physics, University of Hamburg, 
Luruper Chaussee 149, 22761 Hamburg Germany} 
\author{P. Schmelcher}
\affiliation{Center for Optical Quantum Technologies, Department of Physics, University of Hamburg, Luruper Chaussee 149, 22761 Hamburg Germany} 
\affiliation{The Hamburg Centre for Ultrafast Imaging, University of Hamburg Luruper Chaussee 149, 22761 Hamburg, Germany}

\date{\today}

\begin{abstract}
    We study the correlated dynamics of few interacting bosonic impurities immersed in a one-dimensional harmonically trapped bosonic environment. The mixture is exposed to a time-dependent impurity-medium interaction pulse moving it across the relevant phase separation boundary. For modulation frequencies smaller than the trapping one, the system successively transits through the miscible/immiscible phases according to the driving of the impurity-medium interactions. For strong modulations, and driving from the miscible to the immiscible regime, a significant fraction of the impurities is expelled to the edges of the bath. They exhibit a strong localization behavior and tend to equilibrate. Following the reverse driving protocol, the impurities perform a breathing motion while featuring a two-body clustering and the bath is split into two incoherent parts. Interestingly, in both driving scenarios, dark-bright solitons are nucleated in the absence of correlations. A localization of the impurities around the trap center for weak impurity-impurity repulsions is revealed, which subsequently disperse into the bath for increasing interactions.
\end{abstract}

\maketitle
	
	\section{Introduction}
	
	Ultracold atoms serve as an excellent platform to monitor the non-equilibrium dynamics of quantum many-body (MB) systems \cite{Zwerger}, due to the extraordinary experimental tunability of their intrinsic parameters. For instance, the interparticle interactions can be adjusted by means of Feshbach \cite{Fesh1,Fesh2} or confinement induced resonances \cite{CIR,CIR2,CIR3}, and it is possible to realize systems of different dimensionality with arbitrarily shaped trapping potentials \cite{Gorlitz}. Moreover, remarkable progress has been achieved in realizing multicomponent quantum gases \cite{Modugno,Myatt,Egorov,Pilch,Minardi}. A particular focus has been placed on mobile impurities immersed in a MB environment which are consecutively dressed with the collective excitations of the latter thereby forming quasiparticles \cite{Zwierlein,Kohstall,Bruun,Arlt,Jin}. The stationary properties of impurity atoms in a Bose or a Fermi medium \cite{Ardila,Hammer,Guenther,Tempere,Giorgini,Grusdt}, such as their effective mass \cite{Tempere,Grusdt,Ardila2}, lifetime \cite{Kohstall,Jin} and induced interactions \cite{Volosniev,Jie,Amir} have been extensively studied. Recently the emergent dynamics of these settings has been investigated \cite{Widera,Robinson,Schmidt,Giamarchi,Fukuhara,Catastrophe,Knap}, e.g. by dragging impurities in a MB environment \cite{Diss,Coalson}, quenching the impurity-medium interaction strength \cite{Catastrophe,Mista3}, and modifying the external potential experienced by the impurities \cite{Ignuscio,Koushik}. The aforementioned quench protocols have led to dynamical phenomena such as entropy exchange processes between the impurity and the bath \cite{Ignuscio}, dissipative motion of impurities inside Bose-Einstein condensates (BECs) \cite{Diss,Giamarchi}, slow relaxation dynamics \cite{Coalson,Lausch}, the breakdown of the quasiparticle picture for near resonant impurity-bath interactions \cite{Robens}, and the emergence of temporal orthogonality catastrophe phenomena \cite{Knap,Catastrophe}.

	Independently and in a completely different context, non-equilibrium periodic driving protocols of the involved scattering lengths or the trapping potential have been utilized in order to generate and stabilize non-linear excitations such as solitons in one-dimensional (1D) single  \cite{Theocharis,Sakaguchi,Lakshmanan,PanosB}, and two-component BECs \cite{Liu,Rajendran,Kanna}, as well as higher-dimensional settings \cite{Ueda,Caputo}, and also unravel their collisions in a controllable manner \cite{Liu,Kanna}. Interestingly, it has been showcased that the periodic modulation of the interatomic interactions leads to parametrically excited resonant modes and pattern formation in BECs, such as Faraday waves \cite{Pelster,Hulet,Engels,Koushik2,Nicolin,Staliunas,Bychkov}, resembling the response of fluids subjected to a vertical oscillatory force. Moreover, a plethora of additional intriguing phenomena have been exemplified, including the ejection of matter-wave jets in a two-dimensional (2D) cesium BEC \cite{Fireworks,Fireworks2,Fireworks3}, which carry information regarding the phase of the condensate \cite{Fireworks3}, and the emission of correlated atom jets from a bright soliton \cite{Zupanic}.

	Motivated by the above-described phenomena the periodic driving of the impurity-medium interactions provides an interesting avenue to unravel the dynamical response of both subsystems. Given the advances that have been put forward with time-periodic quench protocols, we expect to identify a variety of dynamical response regimes depending on the characteristics of the driving, where for instance spatial localization of the impurities might occur \cite{Umarov}, phase separation phenomena can be manifested, specific patterns can be imprinted in the bath being inherently related to its coherence properties \cite{Li}. For instance, it has been shown that shaking the impurities harmonic trap and depending on the driving frequency leads to intriguing collisional aspects with their host such as a distorted collective dipole motion, their effective trapping or escape from the medium \cite{Koushik}. In this sense, non-linear structures can be spontaneously generated \cite{Grusdt,Lars}, with the time-periodic driving favoring pattern formation in both components of the system \cite{Koushik2}. Additionally the response of few instead of one or two impurities during the dynamics is certainly an interesting aspect. In the present work a pulse of the impurity-medium interactions is employed in order to study the non-equilibrium correlated dynamics of few interacting bosonic impurities embedded in a MB bosonic gas, driving the mixture across its miscibility-immiscibility phase boundary. We track the correlated dynamics of the bosonic mixture by utilizing a variational approach, namely the Multi-Layer Multi-Configuration Time-Dependent Hartree method for Atomic Mixtures (ML-MCTDHX) \cite{ML1,ML2,ML3}.

	First, the particle imbalanced system is driven from the miscible to the immiscible phase and two distinct response regimes are identified. For modulation frequencies smaller than the trapping one, the impurities and the bath successively transit in time through the miscible and immiscible phases according to the temporal driving of their mutual interactions. Turning to larger modulation frequencies, dark-bright (DB) soliton pairs emerge in the absence of correlations forming after half an oscillation period an almost steady bound state around the trap center \cite{PanosB,Lars}. Taking correlations into account, these pairs travel towards the edges of the cloud of their environment, where they remain while oscillating \cite{Mista3,Catastrophe,Diss}. Simultaneously they feature a spatial localization tendency and are two-body correlated between each other. Moreover a density dip (hump) around the trap center is formed for the bath (impurities). Two-body correlations develop for bath particles residing between the two distinct spatial regions separated by the central hump. Employing an effective potential picture \cite{Catastrophe,Diss}, it is found that the impurities reside in a superposition of its lowest-lying eigenstates. In the opposite modulation scenario, where the system is driven to its miscible phase, the two previously mentioned regimes can still be captured. For modulation frequencies larger than the trapping one, oscillating DB solitons emerge within the mean-field (MF) framework \cite{Lars}, which at long evolution times gradually fade away. In sharp contrast within the MB scenario a splitting of the quantum DB soliton pair \cite{Lia3} into two fragments occurs at the initial stages of the dynamics which subsequently fluctuate near the trap center. Accordingly, coherence is almost completely lost for the MB environment. The impurities exhibit a breathing motion, whose frequency is in a good agreement with the predictions of the effective potential, and for longer times they exhibit a two-body clustering \cite{Catastrophe,Bruun,Volosniev}.
	
	Moreover, we inspect the role of impurity-impurity interactions for the cases of two and ten impurity atoms following an impurity-bath interaction pulse from the immiscible to the miscible phase and vice versa. For weak repulsions, the impurities majorly reside in both cases around the trap center, occupying predominantly the ground state of their effective potential \cite{Koushik}. Outer density branches become pronounced only when the particle number or the impurity-impurity repulsion increase. 
	
	This work unfolds as follows. Section \ref{Sec:Framework} introduces our system, describes the employed driving protocol, the used MB ansatz and the observables which will be employed to track the dynamics. Subsequently, in Section \ref{Sec:Immiscible} the non-equilibrium dynamics of the bath-impurity system is explored for a driving from the miscible to the immiscible phase and the reverse scenario is deployed in Section \ref{Sec:Miscible}. Section \ref{Sec:Diff_Imp} elaborates on the dynamical response of the impurities for different particle numbers and impurity-impurity interactions, in both driving scenarios. Finally, in section \ref{Sec:Conclusions} we summarize our main results and suggest possible future extensions of our work. In Appendix \ref{Sec:Appendix_Energy} we briefly discuss the energy exchange processes taking place between the two components, while in Appendix \ref{Sec:Appendix_Breathing}, the breathing frequency of the impurities  is investigated as a function of the modulation frequency within the MF approach for a driving to the miscible phase.

	\section{Theoretical Framework} \label{Sec:Framework}
	
	\subsection{Hamiltonian and driving protocol}

	We consider a particle imbalanced bosonic mixture containing $N_A=100$ atoms forming the environment and $N_B=10$ impurities. The system is mass-balanced, i.e. $M_A=M_B=M$, and it is confined within an elongated harmonic trap of frequency $\omega_A=\omega_B=\omega$. Such a mixture can be realized experimentally, by employing two hyperfine states of $^{87}$Rb, e.g. the $\ket{F=1,m_F=-1}$ for the environment and $\ket{F=2,m_F=1}$ for the impurities \cite{Egorov}. The MB Hamiltonian of this system reads
	
	\begin{gather}
	   \mathcal{H}=\sum_{\sigma=A,B}\sum_{i=1}^{N_{\sigma}}\left[ -\frac{\hbar^2}{2M_{\sigma}}\frac{\partial^2}{\partial(x_i^{\sigma })^2}+\frac{1}{2}M_{\sigma}\omega_{\sigma}^2(x_i^{\sigma})^2\right]+ \nonumber \\ \sum_{\sigma=A,B}g_{\sigma\sigma}\sum_{i<j}\delta(x_i^{\sigma}-x_j^{\sigma})+g_{AB}(t)\sum_{i=1}^{N_A}\sum_{j=1}^{N_B}\delta(x_i^A-x_j^B),
	   \label{Hamilt}
	\end{gather}
	where $g_{\sigma\sigma}$ denotes the two involved intraspecies interaction strengths, $g_{AB}(t)$ is the impurity-medium coupling and $\boldsymbol{x}^{\sigma}=(x^{\sigma}_1,\ldots,x^{\sigma}_{N_{\sigma}})$ are the spatial coordinates of the $\sigma=A,B$ species. The mixture consists of ultracold $^{87}$Rb atoms, and hence its interparticle interactions occur predominantly via $s$-wave scattering \cite{Fesh1}. All the involved effective coupling strengths can be expressed in terms of the corresponding 3D $s$-wave scattering lengths, $a_{\sigma\sigma'}$, and the harmonic oscillator length in the transverse direction $a_{\perp}=\sqrt{\hbar/\mu\omega_{\perp}}$ \cite{CIR,CIR2}. Namely $g_{\sigma\sigma'}=\frac{2\hbar^2 a_{\sigma\sigma'}}{\mu a_{\perp}^2}\left[1-\abs{\zeta(1/2)}\frac{a_{\sigma\sigma'}}{a_{\perp}}\right]^{-1}$, where $\mu=M/2$ is the two-body reduced mass and $\zeta$ is the Riemann zeta function. It is convenient to recast the MB Hamiltonian of Eq. \eqref{Hamilt} in terms of $\hbar \omega_{\perp}$, and in what follows we express all the relevant length, time, and coupling strength scales in units of $\sqrt{\hbar/ M\omega_{\perp}}$, $\omega_{\perp}^{-1}$, and $\sqrt{\hbar^3\omega_{\perp}/M}$ respectively. The trapping frequency is $\omega_A=\omega_B=\omega=0.1$ and the involved intraspecies coupling constants $g_{AA}=1.004$, $g_{BB}=0.9544$ are kept fixed to mimick the experimentally relevant interactions of the above-mentioned $^{87}$Rb, unless it is stated otherwise. To limit the spatial extent of our system, we impose hard-wall boundary conditions at $x=\pm 40$, thereby ensuring that their location does not affect the emergent dynamics. It is also worth commenting that experimentally using for instance $\omega \simeq 2\pi \times 100$ Hz the 1D description holds for $\omega_{\perp} \simeq 2\pi\times 5$ kHz whereas temperature effects are negligible for $k_BT \ll 1.5 \mu\textrm{K}$.
	
	The employed time-periodic pulse protocol involves solely $g_{AB}(t)$, which is sinusoidally modulated in time (for $t \geq 0$), according to
	
	\begin{eqnarray}
	   g_{AB}(t)&=&\left[g^{\textrm{in}}_{AB}+(g^f_{AB}-g^{\textrm{in}}_{AB})\sin^2(\Omega t)\right]\theta\left(\frac{5\pi}{2\Omega}-t\right)+ \nonumber \\& & g^f_{AB}\theta\left(t-\frac{5\pi}{2\Omega}\right),
	   \label{protocol}
	\end{eqnarray}
	for a time span of $T=5\pi/(2\Omega)$, with an amplitude of $|g^f_{AB}-g^{\textrm{in}}_{AB}|$ and frequency $\Omega$ starting from $g_{AB}(0)=g^{\textrm{in}}_{AB}$. Subsequently $g_{AB}(t)$ is held constant at $g_{AB}(t)=g^f_{AB}$, for $t>T$, while $\theta(x)$ is the heaviside function. For $T \to 0$, the driving of $g_{AB}(t)$ occurs only at small time scales, and the protocol effectively reduces to a simple interaction quench, whereas in the limit $T \to \infty$, the bosonic system is subjected to a continuous driving of a small frequency. Furthermore, if $\Omega>\omega$ the system is strongly driven, whereas for $\Omega<\omega$, the pulse lies in the weak driving regime. 
	
\begin{figure}[t!]
		\centering
		\includegraphics[width=0.5\textwidth]{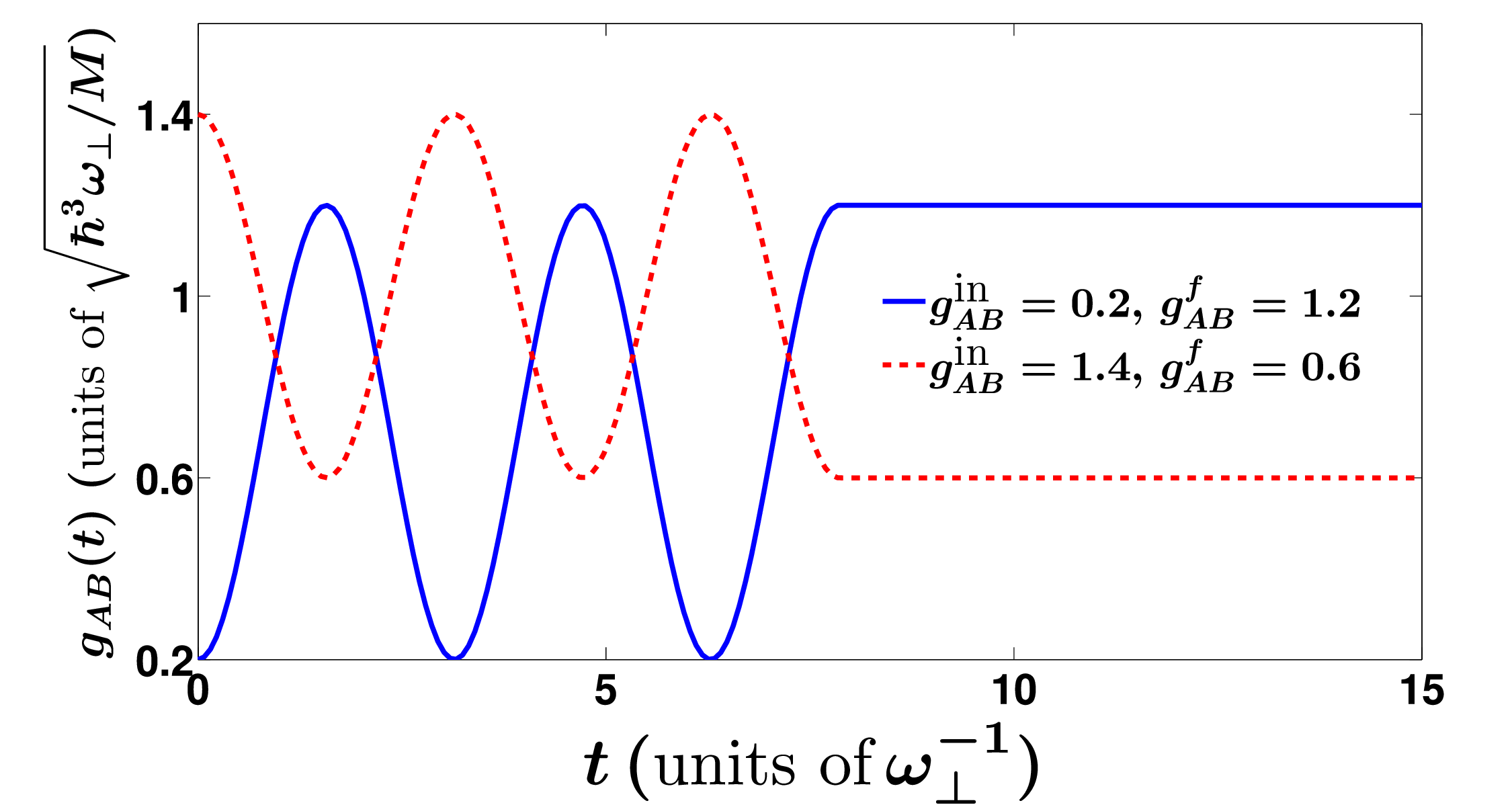}
		\caption{Time-periodic modulation protocol of the impurity-medium interaction strength $g_{AB}(t)$ [Eq. \eqref{protocol}] for the two relevant driving scenarios across the phase separation boundary, namely starting from $g^{\textrm{in}}_{AB}=0.2$ to $g^f_{AB}=1.2$ and from $g^{\textrm{in}}_{AB}=1.4$ to $g^f_{AB}=0.6$.}
		\label{Fig:Quench}
	\end{figure}

	The considered interaction pulse is schematically shown in Fig. \ref{Fig:Quench} when crossing the miscible to the immiscible phase and vice versa. In the following, we will consider two interaction pulse scenarios, both of them driving the impurity-medium interaction strength across the phase separation boundary. For this reason we choose a fixed driving amplitude, namely $|g^f_{AB}-g^{\textrm{in}}_{AB}|$=1. Naturally, a larger driving amplitude crossing the phase separation boundary leads to the same behavior as below, while a smaller amplitude which does not cross the relevant threshold is another interesting case which we do not address in the present work. Recall that phase separation occurs whenever $g_{AB}>\sqrt{g_{AA}g_{BB}}$, a condition that is also adequate in the trapped scenario and then the wavefunctions of the two species have minimal spatial overlap \cite{Ao,Timmermans}. In our case, the threshold takes place at $g_{AB}=0.9789$. In Sec. \ref{Sec:Immiscible}, the dynamics is explored as the impurity-medium coupling strength is driven according to Eq. \eqref{protocol} to the immiscible phase ($g^f_{AB}=1.2$) starting from the system's ground state in the miscible regime, characterized by $g^{\textrm{in}}_{AB}=0.2$. Subsequently, in Sec. \ref{Sec:Miscible} the reverse driving scenario is investigated, and in particular $g^{\textrm{in}}_{AB}=1.4$ with the system being initialized in its ground state is modulated to $g^f_{AB}=0.6$, i.e. towards the miscible regime. More precisely, we aim to understand the driven phase separation process and associated pattern formation in both species depending on the initial phase and the related driving frequency. We shall also briefly comment on the impact of different pulse durations and large modulation frequencies on the driven dynamics. However, a more thorough analysis on this issue, leading possibly to the control of the participating correlations of the emergent patterns is desirable, and is left for future investigations.

	\subsection{Many-body wavefunction ansatz}

	In order to simulate the non-equilibrium driven dynamics of the bosonic mixture, as the impurity-medium interaction strength is sinusoidally modulated, the wavefunction is expanded in a time-dependent and variationally optimized basis, deploying the ML-MCTDHX variational method \cite{ML1,ML2,ML3}. Importantly, this wavefunction ansatz involves two major stages in order to adequately capture the system's correlations. The full wavefunction residing in the composite Hilbert space $\mathcal{H}^{AB}=\mathcal{H}^A\otimes \mathcal{H}^B$, with $\mathcal{H}_A$ and $\mathcal{H}_B$ being the Hilbert spaces of the environment and the impurities respectively, is expressed in the form of a truncated Schmidt decomposition of rank $D$ \cite{Horodecki},
	
	\begin{equation}
	   \Psi_{\textrm{MB}}( \boldsymbol{x}^{A}, \boldsymbol{x}^{B};t)=\sum_{k=1}^{D}\sqrt{\lambda_k(t)}\Psi_k^{A}(\boldsymbol{x}^A;t)\Psi_k^B(\boldsymbol{x}^B;t).
	   \label{Schmidt}
	\end{equation} 
	Here $D\leq \min\left(\dim(\mathcal{H}^A),\dim(\mathcal{H}^B) \right)$ and $\lambda_k(t)$ are the well-known time-dependent Schmidt coefficients. The species functions $\Psi_k^{\sigma}(\boldsymbol{x}^{\sigma};t)$ serve as an orthonormal basis for the $\sigma=A,B$ species and signify the $k$-th mode of entanglement between the two subsystems. If at least two distinct Schmidt coefficients $\lambda_k(t)$ are non-zero, then the two species are entangled since the MB wavefunction $\Psi_{\textrm{MB}}$ of Eq. \eqref{Schmidt} cannot be expressed as a direct product of two states \cite{Horodecki,Mista1} as for instance in the MF case (see below).
	
	At a next step each species function, is accordingly expanded in terms of the permanents of $d_{\sigma}$ time-dependent single-particle functions (SPFs) $\varphi_i$, as follows

	\begin{gather}
	  \Psi_k^{\sigma}(\boldsymbol{x}^{\sigma};t)=\sum_{\substack{n_1,\ldots,n_{d_{\sigma}} \\ \sum n_i=N_{\sigma}}} C_{k, (n_1,\ldots,n_{d_{\sigma}})}(t) \nonumber \\
	  \times \sum_{i=1}^{N_{\sigma}!} \mathcal{P}_i \left[\prod_{j=1}^{n_1}\varphi_1(x^{\sigma}_j;t) \cdots \prod_{j=1}^{n_{d_{\sigma}}} \varphi_{d_{\sigma}}(x^{\sigma}_{n_1+\ldots n_{d_{\sigma-1}}+j};t)\right].
	  \label{SPF}
	\end{gather}  
    Here $C_{k, (n_1,\ldots,n_{d_{\sigma}})}(t)$ denotes the time-dependent expansion coefficients, with $n_i$ being the population of particles occupying the $i$-th SPF, $\varphi_i$. The species function, $\Psi_k^{\sigma}(\boldsymbol{x}^{\sigma};t)$, is thus expanded over all $\binom{N_{\sigma}+d_{\sigma}-1}{d_{\sigma}-1}$ permanents, subject to the constraint $\sum_{i=1}^{d_{\sigma}}n_i=N_{\sigma}$. $\mathcal{P}$ is the permutation operator, exchanging two particles among the SPFs. 
    The above-described variational ansatz captures the presence of interspecies [Eq. (\ref{Schmidt})] and intraspecies [Eq. (\ref{SPF})] correlations, thus testifying the appearance of MB effects that are naturally absent e.g. in a MF treatment.
    
    Employing the Dirac-Frenkel variational principle \cite{Dirac,Frenkel} for the above-described MB variational ansatz [see Eqs. \eqref{Schmidt}, \eqref{SPF}], the ML-MCTDHX equations of motion are derived \cite{ML1}. These equations consist of $D^2$ linear differential equations for $\lambda_k(t)$, which are coupled to $D\left[\sum_{\sigma} \binom{N_{\sigma}+d_{\sigma}-1}{d_{\sigma}-1} \right]$ nonlinear integrodifferential equations for the coefficients $C_{k,(n_1,\ldots,n_{d_{\sigma}})}(t)$, and $(m_A+m_B)$ nonlinear integrodifferential equations for the SPFs. For further details regarding the derivation of the ML-MCTDHX equations of motion, we refer the reader to Refs. \cite{ML1,ML2,ML3}. We should note that employing only a single Schmidt coefficient, $\lambda_1(t)=1$, (i.e. using $D=1$), and one SPF per species, i.e. $d_{\sigma}=1$, results in a product MF state among the two species \cite{Lia3,Mista1}. In this sense all particles of a particular species occupy solely a single wavefunction, namely
    \begin{eqnarray}
        \Psi_{\textrm{MF}}(\boldsymbol{x}^A,\boldsymbol{x}^B;t)&=&\Psi^A_{\textrm{MF}}(\boldsymbol{x}^A;t)\Psi^B_{\textrm{MF}}(\boldsymbol{x}^B;t) \\ \nonumber 
        & &=\prod_{j=1}^{N_A}\varphi^A(x_j^A;t)\prod_{k=1}^{N_B} \varphi^B(x_k^B;t).
        \label{MF_limit}
    \end{eqnarray}
    This yields a set of coupled Gross-Pitaevskii equations for the bosonic mixture \cite{Mista1}. Evidently, within this framework all particle correlations are ignored. Therefore, the comparison of the dynamics of the mixture between the above MF state and the variational ansatz as described by Eqs. \eqref{Schmidt}, \eqref{SPF} sheds light onto the impact of interparticle correlations. Herein we explicate their role in the different driving regions defined with respect to the trap frequency. Note that since the bosonic bath consists of $N_A=100$ atoms its initial (ground state) density profile has a Thomas-Fermi (TF) shape, which is well captured by the MF product state. In this way, at least for the used interaction parameters, the dominant effect of correlations is expected to manifest during the dynamics due to their build-up.

    \subsection{Relevant correlation measures}

    In order to monitor the overall dynamical response of the impurities and their environment as well as to identify their emergent pattern formation, we employ the $\sigma$-species one-body reduced density matrix \cite{Sakmann},
    
    \begin{eqnarray}
       \rho^{(1),\sigma}(x,x';t)&=& N_{\sigma}\int \prod_{j=1}^{N_{\sigma}-1}\, d\tilde{x}_j^{\sigma}\prod_{k=1}^{N_{\bar{\sigma}}} \, dx_k^{\bar{\sigma}} \Psi_{\textrm{MB}}^*(x,\boldsymbol{\tilde{x}}^{\sigma},\boldsymbol{x}^{\bar{\sigma}};t) \nonumber \\ & & \times \Psi_{\textrm{MB}}(x',\boldsymbol{\tilde{x}}^{\sigma},\boldsymbol{x}^{\bar{\sigma}};t),
       \label{Species_density}
    \end{eqnarray}
     where $\sigma=A,B$ and $\boldsymbol{\tilde{x}^{\sigma}}=(x_1^{\sigma},\ldots,x_{N_{\sigma}-1}^{\sigma})$, and $\sigma \neq \bar{\sigma}$. Accordingly, the one-body density of the $\sigma$ species is the diagonal of the one-body reduced density matrix, i.e. $\rho^{(1),\sigma}(x;t)= \rho^{(1),\sigma}(x,x'=x;t)$, and herein it is normalized such that $\int dx\, \rho^{(1),\sigma}(x;t)=N_{\sigma}$. This observable is experimentally accessible via averaging over several single-shot realizations \cite{Jochim,Lia}.  The eigenfunctions of $\rho^{(1),\sigma}(x,x';t)$, $\phi_j^{\sigma}(x;t), \, j=1,\ldots,d_{\sigma}$, are termed natural orbitals \cite{ML1}, and they are normalized to their corresponding eigenvalues dubbed natural populations $n_j^{\sigma}$, i.e. $\int dx\, \abs{\phi_j^{\sigma}(x;t)}^2=n_j^{\sigma}$. Recall that in the MF case $n^A_1=n
    ^B_1=1, \, n^A_{j>1}=n^B_{j>1}=0$, and hence the population of more than a single natural orbital manifests the existence of intraspecies correlations \cite{Mista1}.

	To evince the occurrence of intraspecies correlations of the bath and the impurities we invoke the first-order coherence function  \cite{Mista1,Glauber,Sakmann},
	
	\begin{equation}
	   g^{(1),\sigma}(x,x';t)=\frac{\rho^{(1),\sigma}(x,x';t)}{\sqrt{\rho^{(1),\sigma}(x;t)\rho^{(1),\sigma}(x';t)}}.
	   \label{First_coh}
	\end{equation}
	It takes values in the interval $[0,1]$, and provides a measure of the proximity of the MB state to a MF product state, for a specific set of spatial coordinates, $x$ and $x'$. Two distinct spatial regions are dubbed fully coherent or perfectly incoherent if $\abs{g^{(1),\sigma}(x,x';t)}=1$ or $\abs{g^{(1),\sigma}(x,x';t)}=0$ respectively. When $0<g^{(1),\sigma}<1$, we can infer the presence of intraspecies correlations \cite{Glauber,Mista1}. Recall that for a MF product state [Eq. \eqref{MF_limit}] $g^{(1),\sigma}(x,x';t)=1, \: \forall\,x,x'$ and $ \forall\, t$. 
	
	To capture the appearance of two-body impurity-impurity and bath correlations in a time-resolved manner, we inspect the second-order noise correlation function, $g^{(2),{\sigma\sigma}}(x,x';t)$ \cite{Glauber,Lukin,Mathey}, defined as
	
	\begin{eqnarray}
	  g^{(2),\sigma\sigma}(x,x';t)&=&\rho^{(2),\sigma\sigma}(x,x';t) \nonumber \\
	  & & -\rho^{(1),\sigma}(x;t)\rho^{(1),\sigma}(x';t).
	  \label{Second_coh}
	\end{eqnarray}
	Here, in second quantization $\rho^{(2),\sigma\sigma}(x,x';t)=\braket{\Psi_{\textrm{MB}}(t)|\hat{\Psi}^{\dagger,\sigma}(x')\hat{\Psi}^{\dagger,\sigma}(x)\hat{\Psi}^{\sigma}(x)\hat{\Psi}^{\sigma}(x')|\Psi_{\textrm{MB}}(t)}$ is the diagonal two-body density matrix, and $\hat{\Psi}^{\sigma}(x) [\hat{\Psi}^{\dagger \sigma}(x)]$ is the bosonic operator that annihilates [creates] one particle of species $\sigma$ at position $x$. The diagonal two-body density matrix $ \rho^{(2),\sigma\sigma}(x,x';t)$ provides the probability of simultaneously finding two particles of species $\sigma$ at positions $x$ and $x'$ respectively. Accordingly, the noise correlation function quantifies the presence of two-body correlations between two particles of species $\sigma$ at positions $x$ and $x'$ respectively. The $\sigma$-species MB state is termed two-body correlated [anti-correlated], when $g^{(2),\sigma\sigma}(x,x';t)>0,\, [g^{(2),\sigma\sigma}(x,x';t)<0]$. If $g^{(2),\sigma\sigma}(x,x';t)=0$, then perfect second-order coherence can be inferred. We remark that $ g^{(2),\sigma\sigma}(x,x';t)$ is experimentally probed via \textit{in situ} density-density fluctuation measurements \cite{Tavares}. Moreover, let us note that a MF product state ensures that $g^{(2),\sigma \sigma}(x,x';t)=0 \: \forall \, x,x'$ and $\forall \, t$.
	
	Another important observable, which yields information regarding the spatial extent of each species cloud and thus for its breathing motion, is the position variance \cite{Jens}, 	
	\begin{equation}
	  \braket{(x^{\sigma})^2}=\int \prod_{j=1}^{N_{\sigma}}dx_j^{\sigma} \,\prod_{k=1}^{N_{\bar{\sigma}}} dx_k^{\bar{\sigma}} (\boldsymbol{x}^{\sigma})^2\, \abs{\Psi_{\textrm{MB}}(\boldsymbol{x}^{\sigma},\boldsymbol{x}^{\bar{\sigma}};t)}^2,
	  \label{Position}
	\end{equation}  
	where $\bar{\sigma} \neq \sigma$. This quantity is experimentally accessible via time-of-flight imaging \cite{Ronzheimer}.

	\section{Driven dynamics to the immiscible phase} \label{Sec:Immiscible}

	Below, we discuss the non-equilibrium periodically driven dynamics of the bosonic mixture consisting of a bath with $N_A=100$ atoms and $N_B=10$ impurities. The mixture is initialized in its ground state characterized by $g_{AA}=1.004$, $g_{BB}=0.9544$ and $g^{\textrm{in}}_{AB}=0.2$. Then, the impurity-medium interaction strength is sinusoidally modulated with frequency $\Omega$ for a time span of $T=\frac{5\pi}{2\Omega}$ according to the protocol introduced in Eq. \eqref{protocol}. The modulation drives the mixture into its immiscible phase since the final interaction is $g^f_{AB}=1.2$. To unveil the correlated character of the dynamics we utilize the variational Ans\"atze of Eqs. \eqref{Schmidt}, \eqref{SPF}, and compare with the MF approximation within the ML-MCTDHX framework.

	\begin{figure}[t!]
	\centering
	\hspace*{-0.2cm} 
		\includegraphics[width=0.535 \textwidth,keepaspectratio]{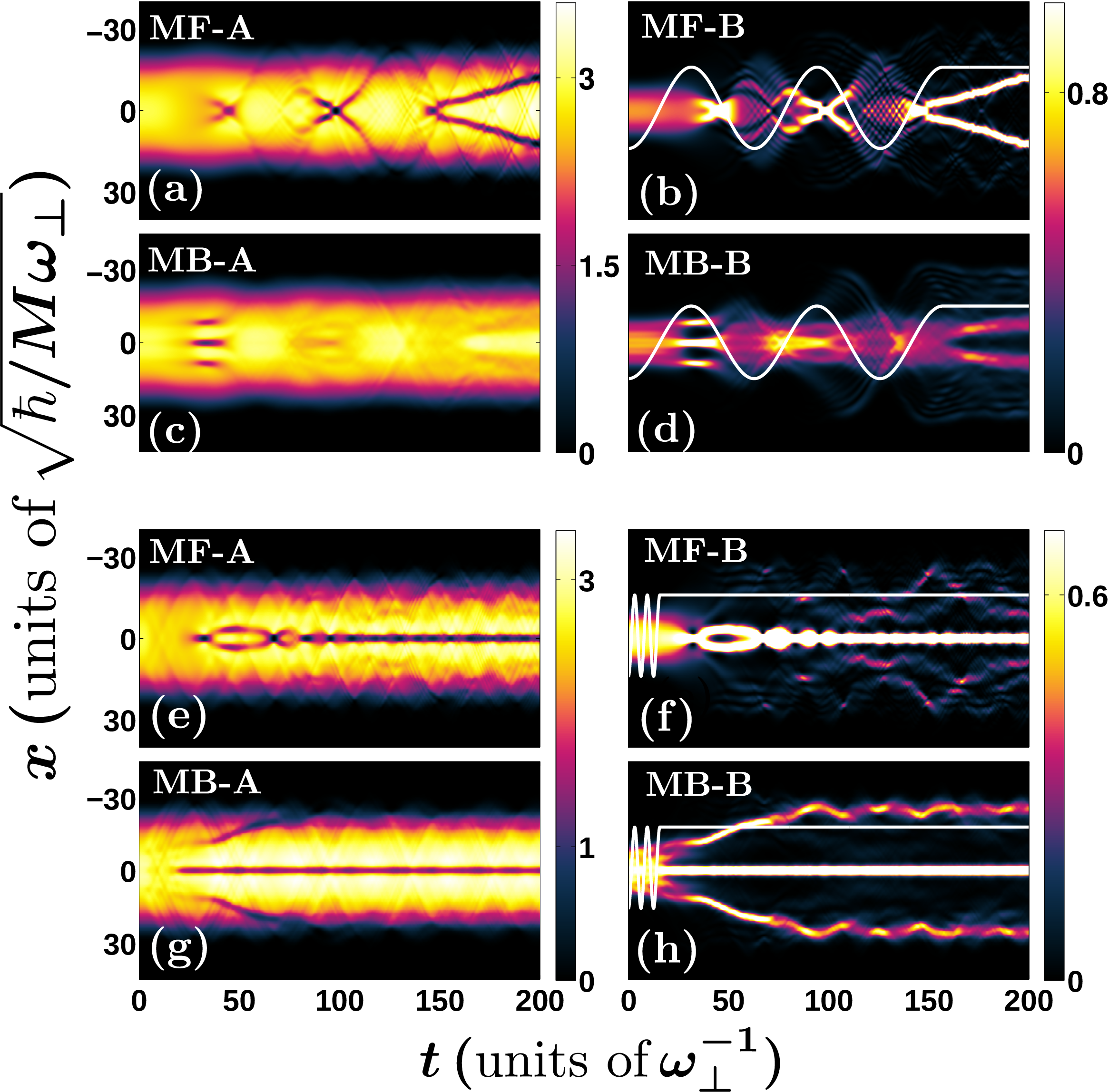}
		\caption{ Spatiotemporal evolution of the one-body density $\rho^{(1),\sigma}(x;t)$ of the impurities (B) and the bosonic bath (A) considering a time-periodic modulation of the impurity-medium coupling [depicted with the solid line in (b), (d), (f), (g)] from $g^{\textrm{in}}_{AB}=0.2$ to $g^f_{AB}=1.2$ for driving frequencies, (a)-(d) $\Omega=0.05$ and (e)-(h) $\Omega=0.5$. The driven dynamics is showcased in the MF approximation in (a), (b), (e), (f) and in the MB approach in (c), (d), (g), (h). The mixture consists of $N_A=100$ and $N_B=10$ particles while it is initialized in its ground state with $g_{AA}=1.004$, $g_{BB}=0.9544$ and $g_{AB}^{\textrm{in}}=1.2$.}
		\label{Fig:Dens_0p2_1p2}
	\end{figure}

	\subsection{One-body density evolution for a pulse with $\Omega<\omega$}
	
\begin{figure}[t!]
		\centering
		\hspace*{-0.25cm}
		\includegraphics[width=0.5 \textwidth,keepaspectratio]{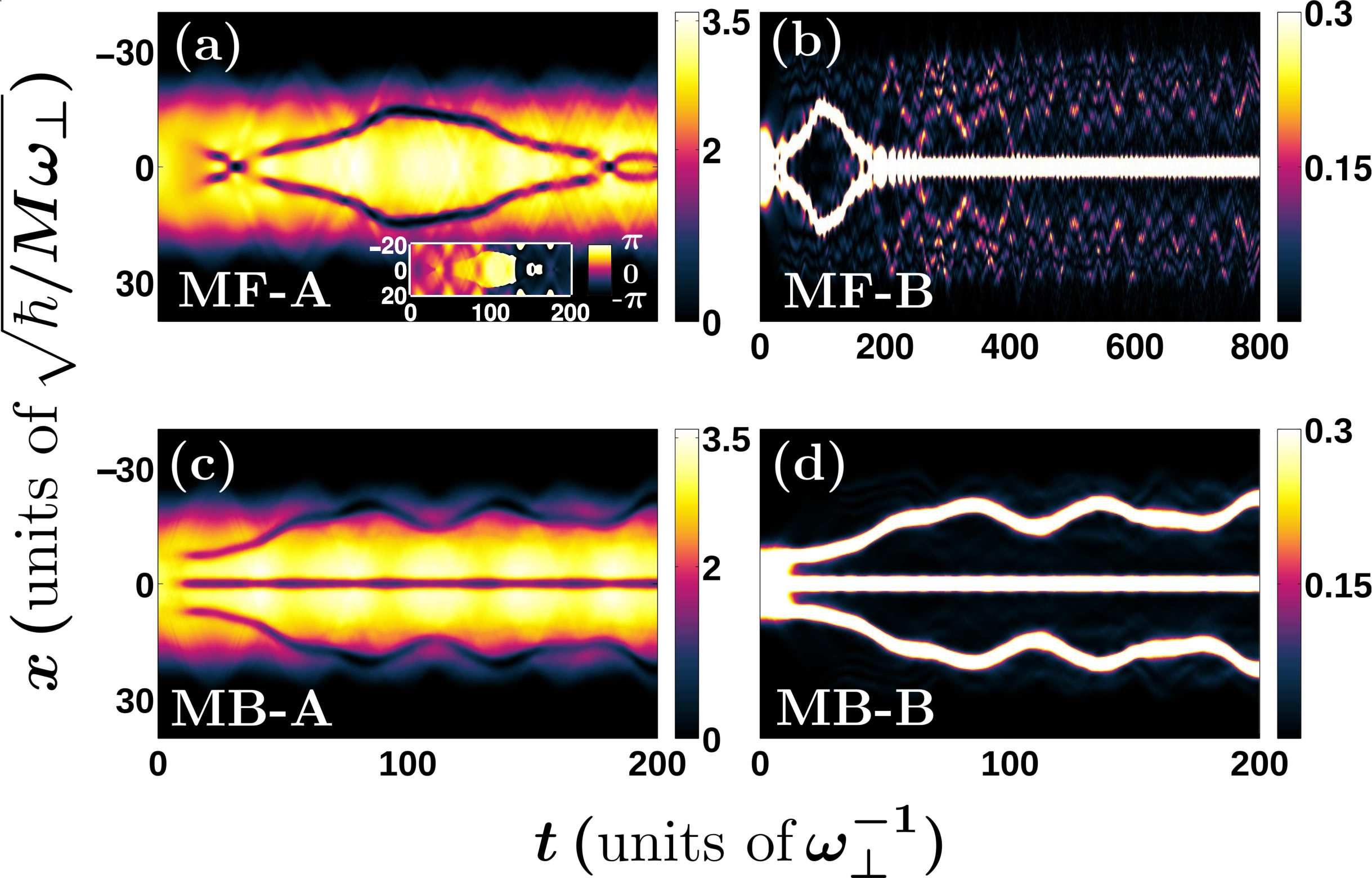}
		\caption{ Temporal evolution of the density $\rho^{(1),\sigma}(x;t)$ of the bath (A) and the impurities (B) following a modulation of the impurity-medium interaction strength from $g_{AB}^{\textrm{in}}=0.2$ to $g_{AB}^f=1.2$ with driving frequency $\Omega=1.5$. The dynamics is compared between (a), (b) the MF approach and (c), (d) the MB method. The inset of (a) illustrates the phase of the bath in the course of the dynamics. The long time-evolution of the impurities for the same modulation and within the MF approximation is presented in (b).}
		\label{Fig:Dens_0p2_1p2_2}
	\end{figure}

	The dynamical response of the bosonic bath and the impurities as captured by the corresponding density evolution is shown in Figs. \ref{Fig:Dens_0p2_1p2}, \ref{Fig:Dens_0p2_1p2_2} for some exemplary modulation frequencies of the impurity-medium pulse protocol of Eq. \eqref{protocol}. As we shall argue below the systems' response is significantly altered for modulation frequencies above ($\Omega>\omega$) and below ($\Omega<\omega$) the trapping frequency. First, we focus on the weak pulse case with $\Omega=0.05<\omega$, presented in Fig. \ref{Fig:Dens_0p2_1p2} (a)-(d). The dynamics within the MF approach [Fig. \ref{Fig:Dens_0p2_1p2} (a), (b)] can be divided into two temporal regimes; one where $g_{AB}(t)$ is modulated across the miscibility threshold (which occurs here at $g_{AB}=0.9789$) for $t\lesssim 157$ and the other for $t>157$, where $g^f_{AB}=1.2$ is constant and the mixture lies in its immiscible phase. In the first regime the impurities and the bath develop simultaneously density humps and dips respectively when $g_{AB}(t)>1$ i.e. within the immiscible phase, see e.g. Fig. \ref{Fig:Dens_0p2_1p2} (b) at $85<t<108$. On the contrary, the impurities feature a diffusive behavior when the bosonic mixture lies in its miscible phase [e.g. at $117<t<138$ in Fig. \ref{Fig:Dens_0p2_1p2} (b)]. As long as the modulation is terminated, i.e. $t>\frac{5\pi}{2\Omega}$, we observe the emission of two counter-propagating impurity density branches, which travel towards the edges of the bath cloud. At later evolution times $t>300$ (not shown here), these branches turn back and collide forming a density dip at the trap center. Accordingly, since $g^f_{AB}=1.2$, the bath density exhibits dips at the locations of the impurities branches as a result of the impurity-medium phase separation \cite{Mista1}, see Fig. \ref{Fig:Dens_0p2_1p2} (a).

	\begin{figure}[t!]
	 	\centering
	 	\hspace*{-0.4cm}
	 	\includegraphics[width=0.52 \textwidth,keepaspectratio]{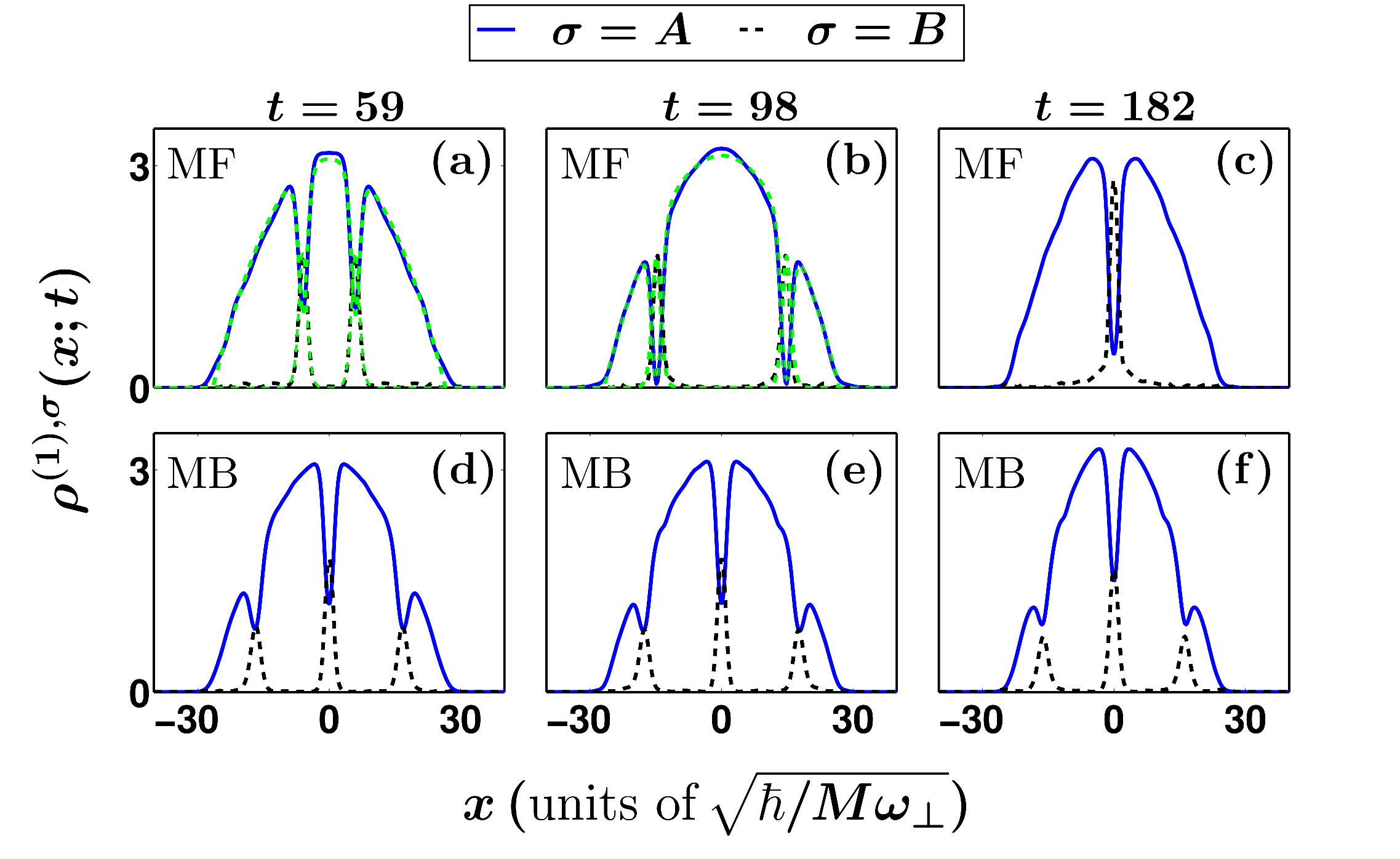}
	 	\caption{Profile snapshots of the one-body density of the impurities (B) and the bath (A) following a time-periodic modulation of the impurity-bath coupling with frequency $\Omega=1.5$ within the (a)-(c) MF and (d)-(f) MB approach (see legend). The dashed green line in (a), (b) represent fittings of the DB soliton ansatz [Eq. \eqref{solitons1}, \eqref{solitons2}] on $\rho^{(1),\sigma}(x;t)$.}
	 	\label{Fig:Snapshots}
	 \end{figure}
	
	In the presence of correlations [Figs. \ref{Fig:Dens_0p2_1p2} (c), (d)], the density of both components exhibits the same qualitative behavior as for the MF evolution, but with some differences which are mainly manifested at later evolution times $(t>40)$. Density humps and dips form on top of the density profiles of the impurities and the bath respectively. These structures differ in their number, position and amplitude from the ones identified within the MF approach as can be seen by comparing Fig. \ref{Fig:Dens_0p2_1p2} (a) and (c) as well as Fig. \ref{Fig:Dens_0p2_1p2} (b) and (d) e.g. at $t\simeq 100$.

   More precisely the aforementioned dips and humps present in the MF scenario [see e.g. Fig. \ref{Fig:Dens_0p2_1p2} (a), (b) at $t=98$] resemble the formation of DB solitons in binary mixtures, where the bright solitons are effectively trapped by the dark ones building upon the bath density \cite{Kevrekidis,Yan,Busch,Middelkamp,Alvarez,Alejandro}. To further support this argument, we perform a fit on the densities of both species at $t=98$ with $\Omega=0.05$, i.e. in the immiscible phase, using the exact single DB soliton wavefunction in the limit where all interactions among and within the species are equal, i.e. the so-called Manakov limit \cite{Yan,Lia2}. The corresponding ansatz for the dark soliton reads
	 
	 \begin{equation}
	     \Psi_{DS}^{\pm}(x,t)=\cos\varphi \tanh \left(d(x\pm x_0(t)) \right)+i\sin\varphi, 
	     \label{solitons1}
	 \end{equation}
	 while for the bright component it has the following form
	 \begin{equation}
	     \Psi_{BS}^{\pm}(x,t)=B \textrm{sech}\left(d(x\pm x_0(t))\right)e^{ikx+i\theta(t)}.
	     \label{solitons2}
	 \end{equation}
    In these expressions $\pm x_0(t)$ are the positions of the dark and bright solitons, $\cos \varphi$ and $B$ denote the amplitudes of dark and bright entities respectively, whereas $d$ is their common inverse width. Moreover, $\sin\varphi$ denotes the dark soliton's velocity, $k=d\tan \varphi$ is the constant wave number of the bright soliton and $\theta(t)$ is its phase. For the fitting of these waveforms to our data we employ $\rho^{(1),A}(x;t)=Q(R^2-x^2)\theta(R^2-x^2)\abs{\Psi^+_{DS}(x,t)}^2\abs{\Psi^-_{DS}(x,t)}^2$ for the bosonic medium, where we assume that the dark solitons are formed on top of a TF profile and $\rho^{(1),B}(x;t)=\abs{\Psi^+_{BS}(x;t)}^2+\abs{\Psi^-_{BS}(x,t)}^2$ for the impurity subsystem. The agreement between the theoretical ansatz and the MF calculations at $t=98$ is adequate, having a standard deviation of the order of $0.03842$ for the dark soliton fit and $0.0847$ for the bright component. At later evolution times $t>150$ [Fig. \ref{Fig:Dens_0p2_1p2} (a), (b)], the density profiles are again reminiscent of DB solitons, however on top of a distorted TF background.
    
    Interestingly, after the termination of the modulation, the density humps (dips) building on top of the density of the impurities (bath) in the MB case [Fig. \ref{Fig:Dens_0p2_1p2} (c), (d)] are less pronounced than the corresponding ones within the MF approach [Fig. \ref{Fig:Dens_0p2_1p2} (a), (b)]. A similar effect, induced by MB correlations has been reported in the case of quantum DB solitons imprinted on BECs, where depleted atoms fill the notch of the dark soliton \cite{Delande,Dziarmaga,Dziarmaga2,Dziarmaga3,Kronke,Burger}.

    \subsection{Density evolution for modulations characterized by $\Omega>\omega$}

    As the modulation frequency becomes larger than the trapping one, the patterns appearing in the one-body density of each component are significantly altered compared to the $\Omega\leq\omega$ case. Characteristic case examples are showcased in Figs. \ref{Fig:Dens_0p2_1p2} (e)-(h) and Fig. \ref{Fig:Dens_0p2_1p2_2} for $\Omega=0.5$ and $1.5$, respectively. This difference to the $\Omega\leq\omega$ scenario is in part due to the fact that the modulation of $g_{AB}(t)$ occurs at very short timescales and the system can not adjust to its very fast external perturbation. Indeed, during the modulation, e.g. until $t \lesssim 15$ in Fig. \ref{Fig:Dens_0p2_1p2} (e)-(h) and $t \lesssim 5$ in Fig. \ref{Fig:Dens_0p2_1p2_2}, $\rho^{(1),\sigma}(x;t)$ exhibits a weak amplitude expansion compared to $\rho^{(1),\sigma}(x;0)$. The magnitude of this expansion is of the order of $3 \%$ and $9\%$ for the bath and the impurities respectively for $\Omega=1.5$ which is in sharp contrast to the $\Omega=0.05$ case [Fig. \ref{Fig:Dens_0p2_1p2} (a)-(d)].

    In particular, within the MF approach and for $\Omega=0.5$ [Figs. \ref{Fig:Dens_0p2_1p2} (e), (f)] a central density hump forms on top of $\rho^{(1),B}$ at the initial stages of the dynamics ($0<t<20$) and subsequently ($t>34$) splits into two density branches, which later on ($t>66$) merge into a central branch propagating undistorted for long evolution times. During the latter process small density portions are emitted travelling towards the edges of the cloud of the bath and back to the trap center. As a consequence of the underlying phase separation mechanism, $\rho^{(1),A}$ displays density dips at the very same positions where the impurities density branches appear [Fig. \ref{Fig:Dens_0p2_1p2} (e)]. Turning to a larger driving frequency [Fig. \ref{Fig:Dens_0p2_1p2_2} (b), $\Omega=1.5$], the impurities density exhibits a two hump structure after $t \gtrsim 12$, which subsequently collide around $t \approx 33$, and afterwards again split moving towards the edges of the bath. These branches collide again at a much later time instant [$t\simeq 180$ in Fig. \ref{Fig:Dens_0p2_1p2_2} (b) and Fig. \ref{Fig:Snapshots} (c)]. In this case a significant portion of energy is pumped into the system and thus both species gain more energy from the modulation compared to the $\Omega=0.5$ scenario, resulting in a larger amount of excitations [see also Appendix \ref{Sec:Appendix_Energy}]. As a consequence, for instance the impurities have enough energy to reach the edges of the bath before colliding again at the trap center. It is also worth mentioning that in the long time dynamics, see Fig. \ref{Fig:Dens_0p2_1p2_2} (b), the impurities density branches merge after $t \simeq 180$ into a single central hump, which stays unperturbed throughout evolution. This hump comes along with small fluctuating emitted density branches, which diffuse within the background density of the medium [hardly visible in Fig. \ref{Fig:Dens_0p2_1p2_2} (b)]. We note that the aforementioned merger of the impurities density branches occurs at earlier times accompanied by a larger amount of excitations in the BEC background as the pulse duration increases since more energy is pumped into the system.
    
    \begin{figure}[t!]
    	\centering
    	\hspace*{-0.1cm}
    	\includegraphics[width=0.52 \textwidth,keepaspectratio]{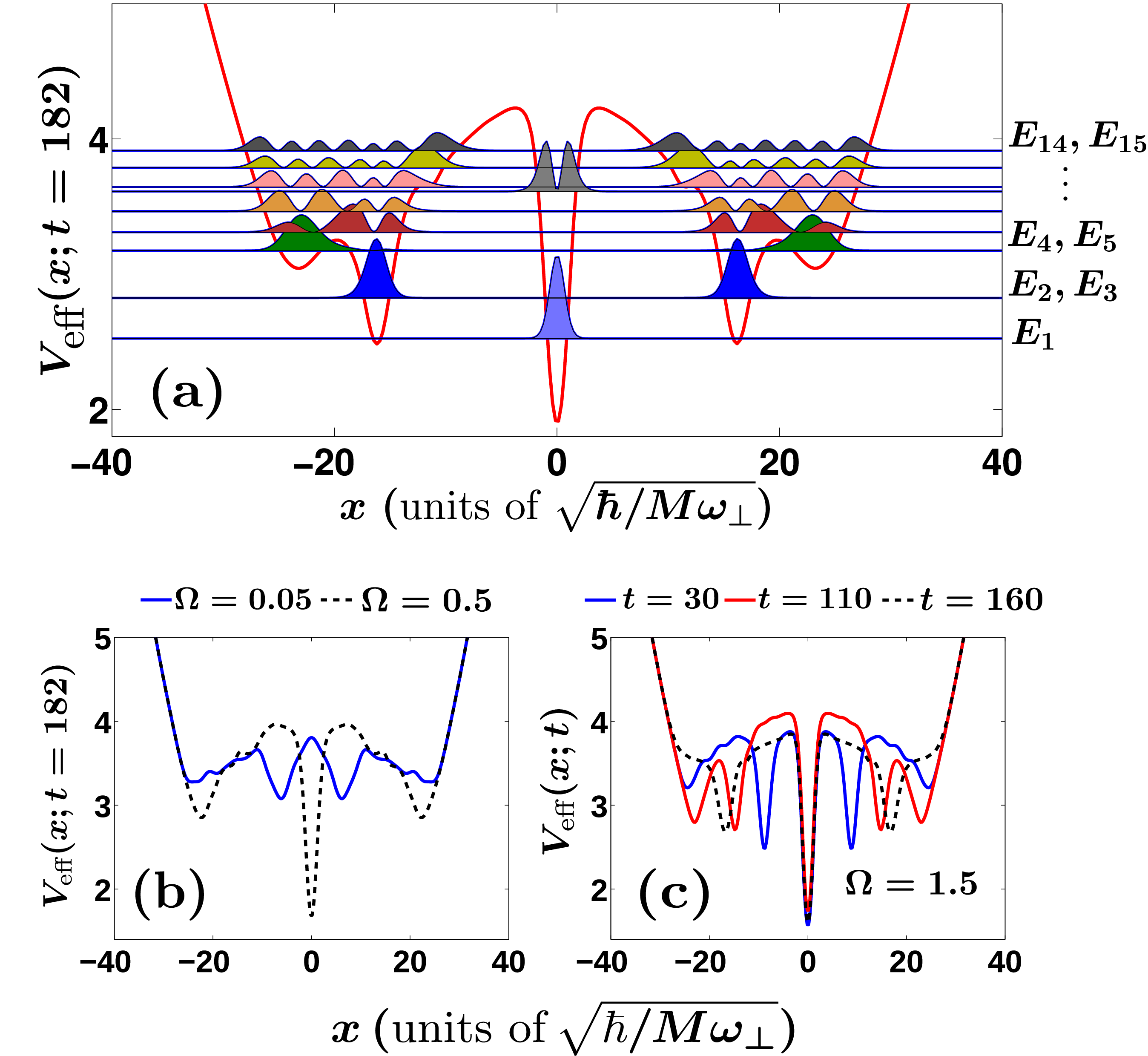}
    	\caption{(a) Instantaneous effective potential at $t=182$ for a modulation frequency $\Omega=1.5$. On top of $V_{\textrm{eff}}(x;t)$ its single-particle eigenstates are depicted together with their energies. (b) The effective potential for other driving frequencies (see legends) at $t=182$. (c) $V_{\textrm{eff}}(x;t)$ at distinct time-instants for $\Omega=1.5$. The effective potential is measured in units of $\hbar \omega_{\perp}$.}
    	\label{Fig:Effective_pot}
    \end{figure}
    
    The above-described density dips [peaks] displayed in the bath cloud [impurities], see Figs. \ref{Fig:Dens_0p2_1p2_2} (a) and (b) are once more reminiscent of the dynamical formation of DB solitons. Indeed, the DB soliton waveform [Eqs. \eqref{solitons1}, \eqref{solitons2}] serves as a good candidate to the density profiles of both components [see the dashed green lines in Figs. \ref{Fig:Snapshots} (a), (b)] with the corresponding fit exhibiting a standard deviation of the order of 0.07. Moreover, the spatiotemporal evolution of the phase of the MF bath wavefunction displays jumps being multiples of $\pi$, as can be seen in the inset of Fig. \ref{Fig:Dens_0p2_1p2_2} (a) at $t>100$, which is a characteristic feature of dark solitons.
    
    The inclusion of correlations leads to a drastically different time-evolution of both the impurities and the medium than in the MF approach see Figs. \ref{Fig:Dens_0p2_1p2} (g), (h) and Figs. \ref{Fig:Dens_0p2_1p2_2} (c), (d) for $t>5$. Indeed the impurities density branches formed after the modulation, move to the edges of the bath cloud, where they perform a weak amplitude oscillatory motion having an equilibration tendency. We remark that an analogous response of the impurities has been demonstrated in the impurity-medium interaction quench dynamics of two spin-polarized fermions inside a Bose gas \cite{Lars}. Also similar dynamical phase separation phenomena have been shown to occur for strong impurity-medium interactions signifying temporal orthogonality catastrophe phenomena of the Bose-polaron \cite{Catastrophe,Diss,Mista3}. Moreover, there is a central density hump (dip) in the density of the impurities (bath). The major difference between $\Omega=0.5$ and $\Omega=1.5$ within the MB framework is that the density branches in the former case [Fig. \ref{Fig:Dens_0p2_1p2} (h)] reach the edges of the bosonic medium, their amplitude decreases and they undergo smaller amplitude oscillations than the ones for $\Omega=1.5$ [Fig. \ref{Fig:Dens_0p2_1p2_2} (d)]. Inspecting the instantaneous MB density profiles $\rho^{(1),\sigma}(x;t)$ [Figs. \ref{Fig:Snapshots} (d)-(f)] when $\Omega=1.5$, we observe that the side humps [dips] for the impurities [bath] appearing around $x \simeq 16$ have a smaller amplitude and are displaced with respect to the ones emerging within the MF approach. Another difference occurring between the MF and the MB evolution is the formation of a central density hump (dip) for the impurities (bath), in addition to the side humps and dips when correlations are present, as can be seen in Figs. \ref{Fig:Snapshots} (d)-(f). We should note that as we increase the modulation frequency $\Omega$ and thus tending to the abrupt quench scenario [Eq. (\ref{protocol})], a similar dynamics to the one illustrated in Fig. \ref{Fig:Dens_0p2_1p2_2} (c), (d) for $\Omega=1.5$ takes place for both components. The most notable difference is that the separation of the outer $\rho^{(1),B}(x;t)$ branches becomes slightly larger.
    
    An intuitive understanding of the response of the impurities is provided by constructing an effective potential picture \cite{Catastrophe,Theel}. The latter is derived from the impurities external trapping potential, and the one-body density of their bosonic medium \cite{Catastrophe,Ferrier,Theel}, namely
    
    \begin{equation}
       V_{\textrm{eff}}(x;t)=\frac{1}{2}M \omega^2x^2+g_{AB}(t)\rho^{(1),A}(x;t).
       \label{Effective_potential} 
    \end{equation}
    Evidently $V_{\textrm{eff}}(x;t)$ is a time-dependent single-particle potential, which is in general different from the external harmonic trap due to its second contribution accounting for the bath and the impurity-medium interactions. 
    Before proceeding we should clarify that $V_{\textrm{eff}}(x;t)$ is not able to account for impurity-medium correlations and as a consequence it does not provide insights into e.g. two-body mechanisms such as the emergent impurity-impurity induced correlations as has been argued in Refs. \cite{Catastrophe,Mista3}. Of course, all these processes are naturally included within our MB treatment performed within the ML-MCTDHX approach.
    
    For instance, $V_{\textrm{eff}}(x;t)$ for $\Omega=1.5$ features a deep central well, present throughout the evolution, and additional shallower side wells whose depths and positions change with time [Fig. \ref{Fig:Effective_pot} (a) and (c)]. These potential wells are a manifestation of the density dips of the bath displayed for instance in Fig. \ref{Fig:Dens_0p2_1p2_2} (c). Even though the effective potential yields a single-particle picture, one can readily see that $\rho^{(1),B}(x;t=182)$ in Fig. \ref{Fig:Dens_0p2_1p2_2} (d) mainly resides in a superposition of the ground and the first two excited states of $V_{\textrm{eff}}(x;t)$, with corresponding participation weights $41 \%$ and $23.6 \%$, $23.6 \%$ respectively. There are also additional density modulations, [hardly visible in Fig. \ref{Fig:Dens_0p2_1p2_2} (d)], which suggest the occupation of higher-lying excited states as well, with a small non-vanishing population up to the $18^{\textrm{th}}$ excited state. At other time instants [Fig. \ref{Fig:Effective_pot} (c)], the depth of the central well of $V_{\textrm{eff}}(x;t)$ changes slightly with time and the outer wells are displaced, accounting thus for the oscillations of the outer density branches shown in Fig. \ref{Fig:Dens_0p2_1p2_2} (d). Apart from the aforementioned undulations during the time-evolution, the effective potential changes also with respect to the driving frequency $\Omega$, since the density profile of the medium is accordingly modified. For instance, the central potential dip is absent in the case of $\Omega=0.05$ [Fig. \ref{Fig:Effective_pot} (b)] and the effective potential displays a double-well structure accounting for the impurities density peaks [Fig. \ref{Fig:Dens_0p2_1p2} (d)]. For $\Omega=0.5$, the central density dip of $\rho^{(1),A}$ forms, which mainly attracts the impurity atoms since its depth is larger than the one of the outer wells.

    \subsection{Correlation dynamics and impurities anti-bunching}

    \begin{figure}[t!]
    	\centering
    	\hspace*{-0.2cm}
    	\includegraphics[width=0.52 \textwidth,keepaspectratio]{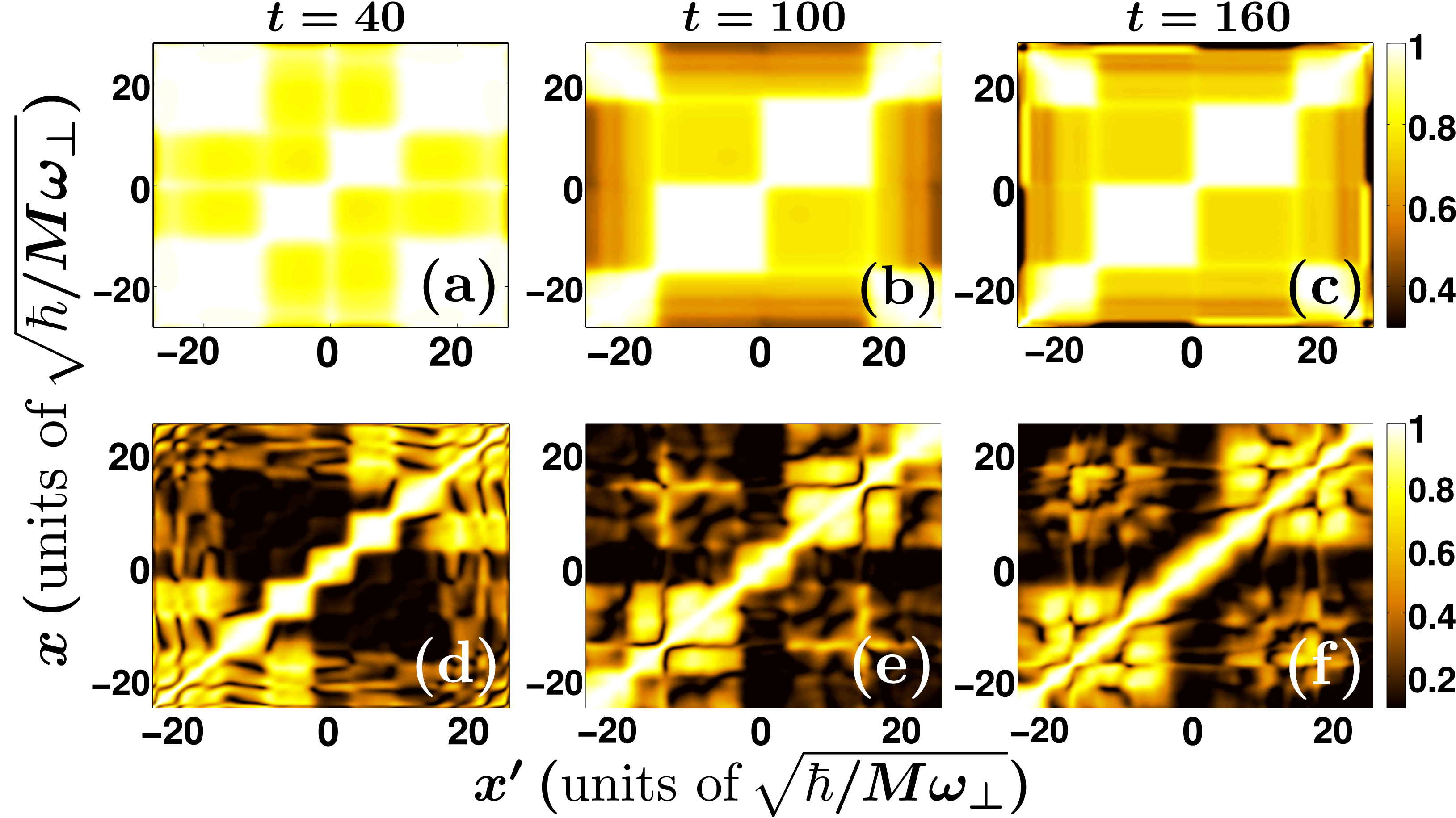}
    	\caption{Snapshots of the first-order coherence function of the (a)-(c) bath $g^{(1),A}$ and the (d)-(f) impurities $g^{(1),B}$ for a driving frequency $\Omega=0.5$. Other system parameters are the same as in Fig. \ref{Fig:Dens_0p2_1p2} (g), (h).}
    	\label{Fig:Coh}
    \end{figure}

    Having explicated the imprint of correlations in the density evolution of both the impurities and the bath we subsequently inspect the first- and second-order correlation functions as introduced in Eqs. \eqref{First_coh} and \eqref{Second_coh} respectively. In this way, we will be able to demonstrate from $g^{(1),\sigma}$ the possibly emergent coherence losses of each species when $g^{(1),\sigma}<1$. Along the same lines, utilizing $g^{(2),\sigma\sigma}$ the two-body correlation properties of the impurities and the bath can be identified for $g^{(2),\sigma\sigma}\neq 0$. Initially, the first-order coherence $g^{(1),\sigma}(x,x';t)$ is examined [Eq. \eqref{First_coh}], from which one can infer the proximity of a MB to a MF product state for a specific set of spatial coordinates $x$ and $x'$ at time $t$. Instantaneous profile snapshots of
    $g^{(1),A}(x,x';t)$ and $g^{(1),B}(x,x';t)$ are shown in Fig. \ref{Fig:Coh} exemplarily for $\Omega=0.5$. 
    
    At early evolution times ($t\leq 40$), where the two impurity density branches travel to the edges of the bath cloud [Figs. \ref{Fig:Dens_0p2_1p2} (g), (h)], the BEC background exhibits relatively small coherence losses, see the off-diagonal of $g^{(1),A}$ [Fig. \ref{Fig:Coh} (a)]. Indeed, it appears that the two separate spatial intervals of the medium enclosed by the central and the outer density dips [see Fig. \ref{Fig:Dens_0p2_1p2} (g)], namely $D^+=(0,+10.56)$ and $D^-=(-10.56,0)$ are slightly off-coherent between each other as well as with the regions from the outer density dips until the edges of the bath cloud, see e.g. $g^{(1),A}(6.02,-6.02;t=40)\simeq 0.8$ [Fig. \ref{Fig:Coh} (a)]. At later times, see for instance Figs. \ref{Fig:Coh} (b) and (c), the spatial domains separated by the central dip at $x=0$, namely $D^+=(0,19)$ and $D
   ^-=(-19,0)$, become less coherent with respect to one another and e.g. $g^{(1),A}(12.17,-12.17;t=160)\simeq 0.75$. On the other hand, the two density branches of the impurities [Fig. \ref{Fig:Dens_0p2_1p2} (h)], are entirely non-coherent throughout the time-evolution, see in particular Figs. \ref{Fig:Coh} (d), (e) and (f) where $g^B(x,x'\neq x;t>40)$ is vanishing. Therefore, the impurities develop Mott-like correlations, suggesting their spatial localization tendency in the two separate density branches \cite{Sherson,Mista1}. When the impurity density branches lie at the edges of their background for $t \gtrsim 62$ and are weakly oscillating, a small amount of coherence is restored e.g. $g^{(1),B}(12.17,-12.17;t=100) \simeq 0.4$ between the emitted faint density peaks located in the spatial regions $x \in [9,13], \, x' \in [-13,-9]$ [Fig. \ref{Fig:Dens_0p2_1p2} (h)]. We remark that a similar coherence behavior occurs also for other modulation frequencies larger than the trapping one. For $\Omega<\omega$ the medium remains almost perfectly coherent throughout the time-evolution and the impurities are localized either in $x>0$ or $x<0$.

    \begin{figure}[t!]
    	\centering
    	\hspace*{-0.8cm}
        \includegraphics[width=0.53 \textwidth,keepaspectratio]{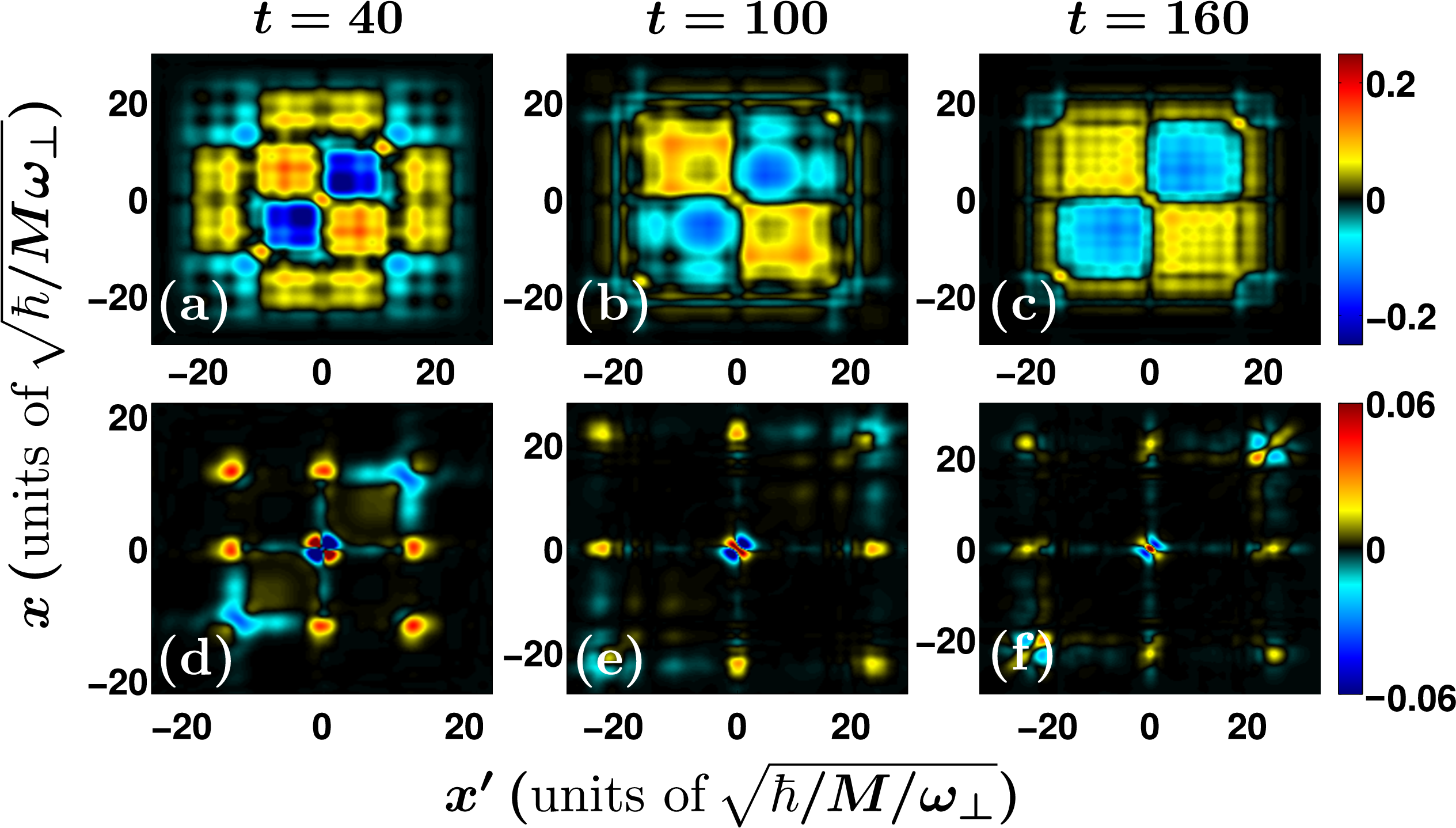}
        \caption{Instantaneous profiles of the second-order noise correlation of (a)-(c) the bath particles $g^{(2),AA}(x,x';t)$ and (d)-(f) the impurities $g^{(2),BB}(x,x';t)$. In all cases the modulation frequency is $\Omega=0.5$, while other system parameters are the same as in Fig. \ref{Fig:Dens_0p2_1p2} (g), (h).}
        \label{Fig:Coh2}
    \end{figure}

    Next, we discuss the two-body correlation characteristics of the impurities and their BEC background, by monitoring $g^{(2),AA}$ and $g^{(2),BB}$ respectively [Fig. \ref{Fig:Coh2}] for $\Omega=0.5$. Focusing on the BEC medium, we observe that for $t \leq100$ where the impurities density humps travel to the edges of the medium cloud [Figs. \ref{Fig:Coh2} (a), (b)], two particles of the environment tend to avoid each other within the two spatial intervals enclosed by the central and outer density dips of $\rho^{(1),A}(x;t)$, i.e. $D_+\simeq(0,20)$ and $D_-\simeq (-20,0)$ [Fig. \ref{Fig:Dens_0p2_1p2} (g)] since $g^{(2),AA}(x,x'=x;t \leq 100)<0$. However, there is an increased probability of finding one of the particles in one of those intervals, e.g. in $D_+$ and the other particle being symmetrically placed with respect to the trap center, e.g. in $D_-$. This behavior persists at later evolution times, as can be seen in Fig. \ref{Fig:Coh2} (b), (c), where two-body correlations build up for particles residing in opposite spatial regions with respect to the trap center, see the anti-diagonal of $g^{(2),AA}(x;t)$. Moreover, a two-body anti-correlation tendency occurs between $D_+$ and $D_-$ since $g^{(2),AA}(x,x'\neq x;t)<0$.

    Turning to the impurities, anticorrelations appear for particles occupying the same position for $t \leq 100$ where the impurities move to the edges of the background cloud [Figs. \ref{Fig:Coh2} (d) and (e)]. However, two particles residing in different density branches [Fig. \ref{Fig:Coh2} (d)] display a correlated character, e.g. $g^{(2),BB}(-10.84,-10.84;40) \simeq -0.0345$ and $g^{(2),BB}(-10.84,10.84;40)\simeq 0.012$. For longer evolution times [Fig. \ref{Fig:Coh2} (f)] two-body correlations build among particles occupying the three distinct density branches, see for instance $g^{(2),BB}(22.61,22.61;t=160)\simeq 0.01$ \cite{Mista1,Jens,Jenny}. In contrast, two particles are anticorrelated when they both lie in the same density hump e.g. the one close to the trap center where $g^{(2),BB}(-0.94,-0.94;160)\simeq -0.05$.

    \section{Driven dynamics to the miscible phase} \label{Sec:Miscible}

    \begin{figure}[t!]
    	\centering
    	\hspace*{-0.4cm} 
    	\includegraphics[width=0.55 \textwidth,keepaspectratio]{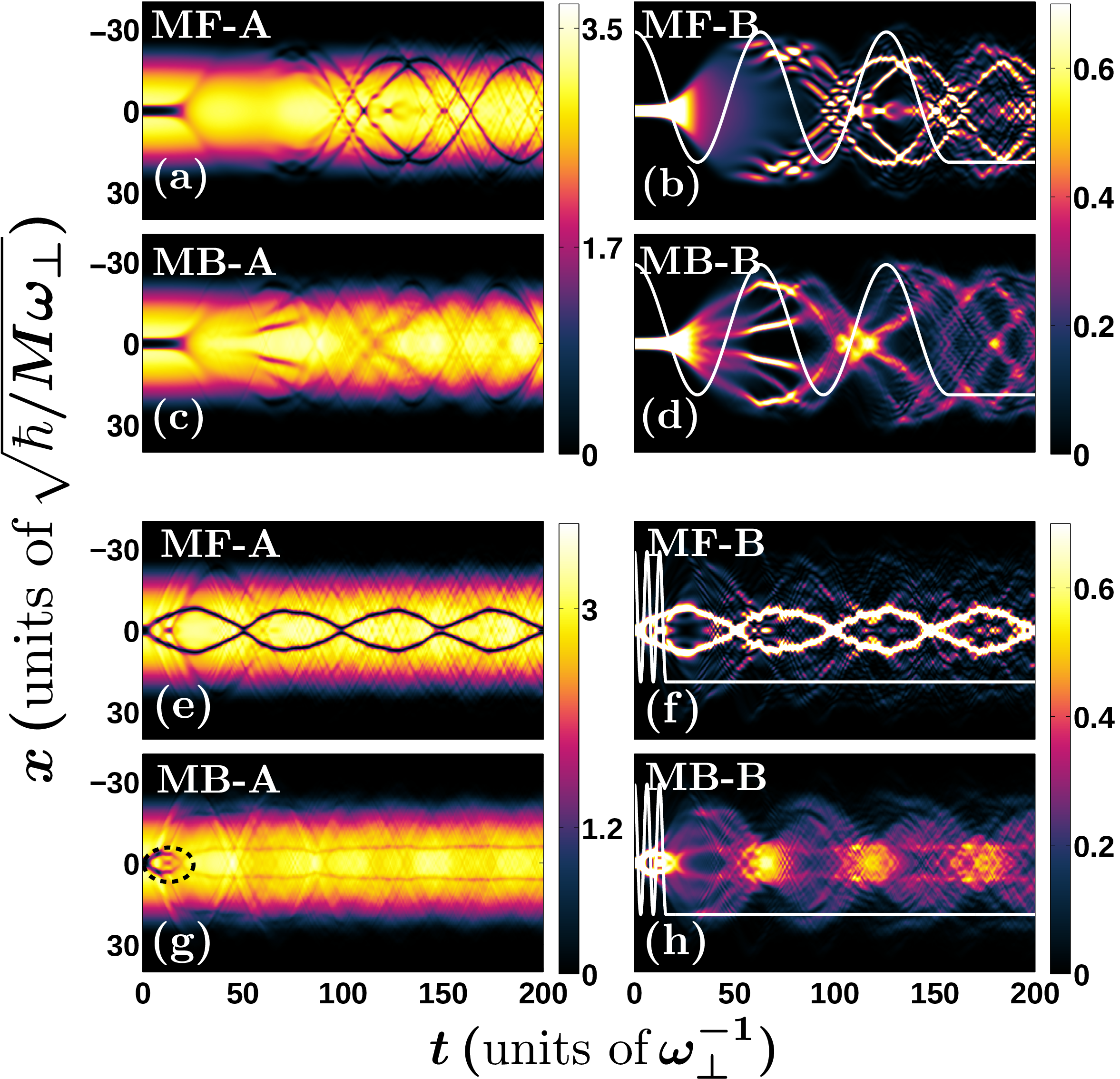}
    	\caption{Time-evolution of the one-body density of (a), (c), (e), (g) the bosonic bath and (b), (d), (f), (h) the impurities following a modulation of the impurities-medium coupling from $g^{\textrm{in}}_{AB}=1.4$ to $g^f_{AB}=0.6$. The modulation is exemplarily depicted with the white solid line in (b), (d), (f), (h). The dynamics is tracked for two driving frequencies, (a)-(d) $\Omega=0.05$ and (e)-(h) $\Omega=0.5$ within the (a), (b), (e), (f) MF approach and (c), (d), (g), (h) the MB method. The mixture comprises of $N_A=100$ bath and $N_B=10$ impurity atoms, characterized initially (ground state) by $g_{AA}=1.004$, $g_{BB}=0.9544$ and $g_{AB}^{\textrm{in}}=1.4$.}
    	\label{Fig:Density_1p4_0p6}
    \end{figure}

  We proceed by analyzing the reverse pulse driving scenario, namely the one where the mixture is driven from the immiscible to the miscible phase, according to the time-dependent protocol of Eq. \eqref{protocol}. More specifically, the mixture is initially prepared in its ground state, characterized by $g_{AA}=1.004$, $g_{BB}=0.9544$ and $g^{\textrm{in}}_{AB}=1.4$. The final impurity-medium interaction strength is $g^f_{AB}=0.6$. As before, the non-equilibrium dynamics is investigated, while it is compared and contrasted between the MF and the MB framework.

  \begin{figure}[t!]
  	\centering
  	\hspace*{-0.65cm} 
  	\includegraphics[width=0.53 \textwidth,keepaspectratio]{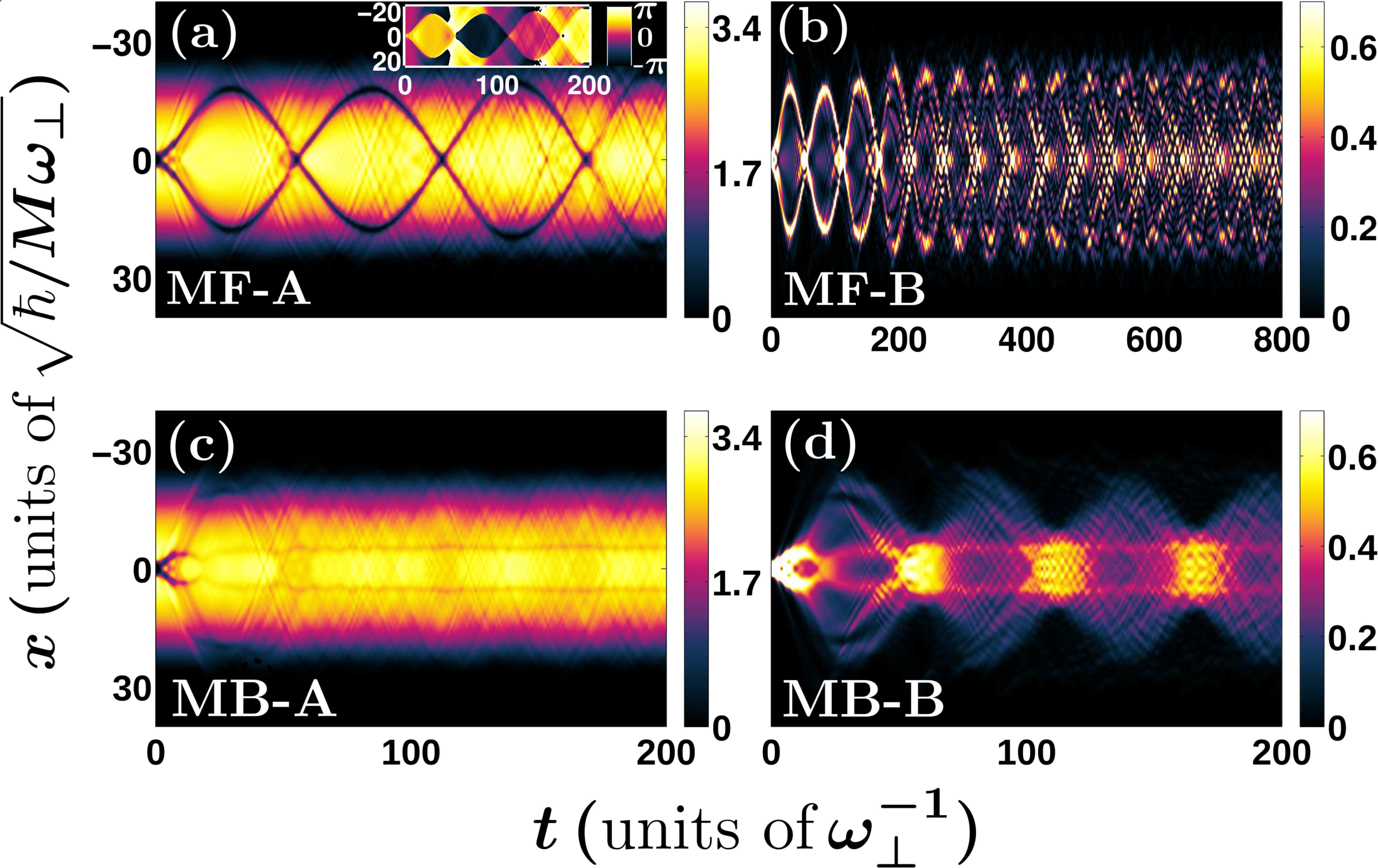}
  	\caption{Spatiotemporal evolution of the one-body density of (a), (c) the bosonic environment and (b), (d) the impurities applying a modulation of the impurity-medium interaction from $g^{\textrm{in}}_{AB}=1.4$ to $g^f_{AB}=0.6$ with frequency $\Omega=1$. The dynamics is displayed both within the (a), (b) MF and (c), (d) MB approaches. The inset of (a) showcases the phase of the bath throughout the time-evolution. The long time-evolution of the impurities within the MF approach is presented in (b).}
  	\label{Fig:Density_1p4_0p6_2}
  \end{figure}

  \subsection{Dynamical response for $\Omega<\omega$}
  
  As explicated in Sec. \ref{Sec:Immiscible}, monitoring the one-body density evolution of the participating components, it is possible to distinguish two driving related response regimes, namely $\Omega<\omega$ and $\Omega> \omega$, see Figs. \ref{Fig:Density_1p4_0p6}, \ref{Fig:Density_1p4_0p6_2} respectively. First, let us focus on the case of $\Omega=0.05<\omega$ [Fig. \ref{Fig:Density_1p4_0p6} (a), (b)], and inspect the dynamical behavior of the bath and the impurities within the MF framework. As it can be readily seen in Fig. \ref{Fig:Density_1p4_0p6} (b), $\rho^{(1),B}(x;t)$ exhibits density humps filling the dips of the bosonic medium, within the time intervals where $g_{AB}>1$, and diffusive patterns as the system is driven to its miscible phase \cite{Koushik}, i.e. $g_{AB}<1$. To facilitate this observation, a white solid line indicating the modulation of the impurity-medium coupling is depicted in Fig. \ref{Fig:Density_1p4_0p6} (b).
  The inclusion of correlations results in a similar dynamical response of the mixture at early evolution times ($t<40$) but subsequently significant alterations take place [Fig. \ref{Fig:Density_1p4_0p6} (c), (d)].

  \begin{figure}[t!]
  	\centering
  	\includegraphics[width=0.51 \textwidth,keepaspectratio]{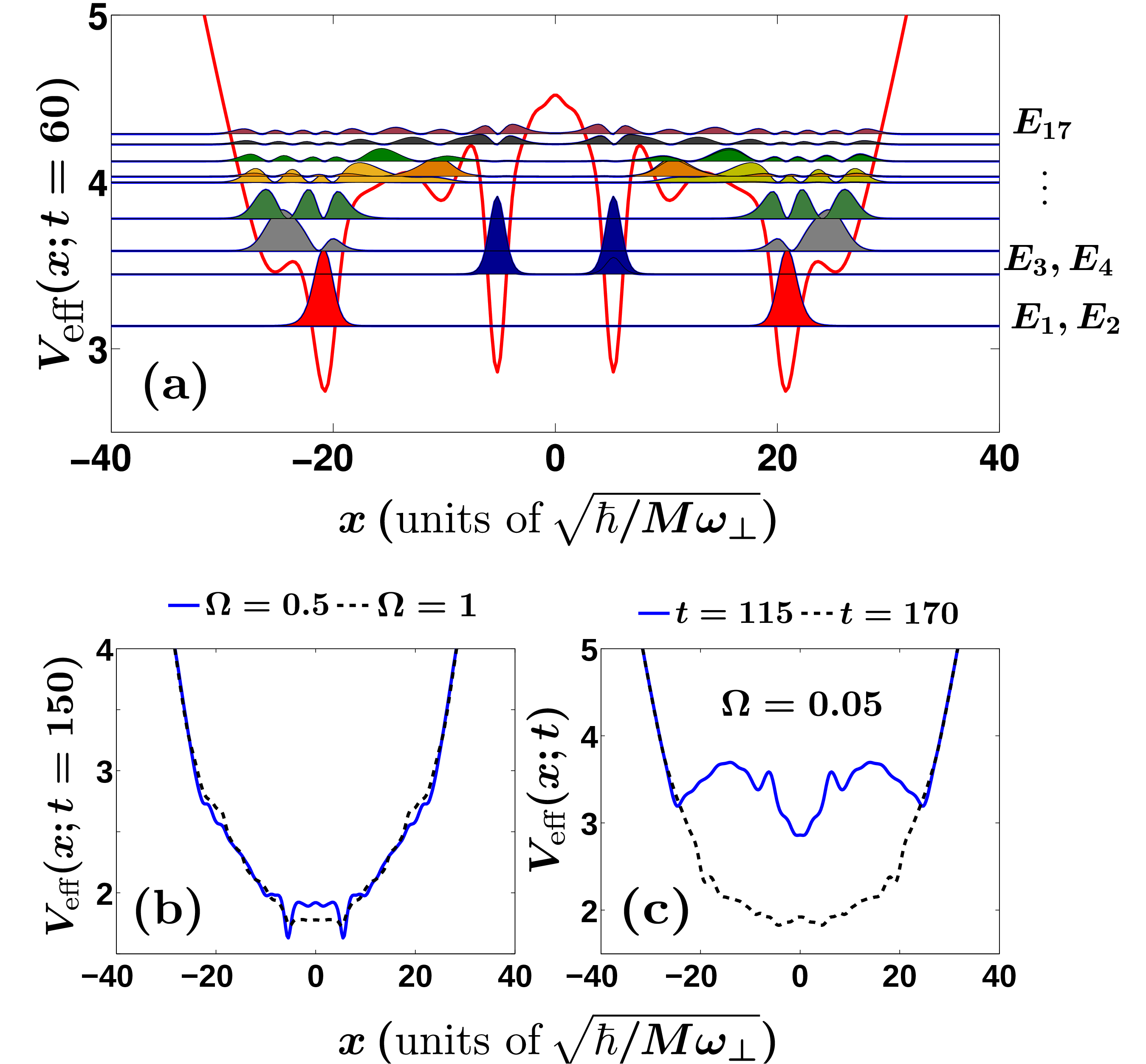}
  	\caption{(a) Instantaneous effective potential at $t=60$ and modulation frequency $\Omega=0.05$. On top of $V_{\textrm{eff}}(x;t)$, its eigenstates are displayed together with their energy. (b) The effective potential for other driving frequencies, (see legends), at $t=150$. (c) $V_{\textrm{eff}}(x;t)$ at other time instants for $\Omega=0.05$. The effective potential is measured in units of $\hbar \omega_{\perp}$.}
  	\label{Fig:Effective_pot2}
  \end{figure}

  Concretely, in the MB approach [Fig. \ref{Fig:Density_1p4_0p6} (c), (d)] a smaller number of generated density dips and humps in $\rho^{(1),A}(x;t)$ and $\rho^{(1),B}(x;t)$ respectively is observed compared to the MF case \cite{Jenny,Hannes,Mista1}, see e.g. $\rho^{(1),B} (x;t)$ at $t\simeq 97$. 
  Later on, the diffusive character of $\rho^{(1),B} (x;t)$ is more pronounced than within the MF treatment, compare for instance Fig. \ref{Fig:Density_1p4_0p6} (b) and (d) around $t\simeq 157$.
  When correlations are present, the impurities are effectively trapped at the density dips developed in the bosonic bath [Fig. \ref{Fig:Density_1p4_0p6} (d)]. The medium thus provides an effective potential experienced by the impurities, and their density profile can be understood by resorting to the potential defined in Eq. \eqref{Effective_potential}. As shown in Fig. \ref{Fig:Effective_pot2} (a), the underlying $V_{\textrm{eff}}(x;t=60)$ features four deep wells, caused by the density profile of the medium. Inspecting the density profile of the impurities [Fig. \ref{Fig:Density_1p4_0p6} (d)] and the form of the effective potential [Fig. \ref{Fig:Effective_pot2} (a)], one can infer that $\rho^{(1),B}(x;t=60)$ mainly resides in a superposition of the four lowest lying eigenstates, $E_1,\ldots E_4$ of $V_{\textrm{eff}}(x;t)$ \cite{Catastrophe}. At later time instants $\rho^{(1),B}(x;t=150)$ presents a diffusive behavior throughout the environment, with a small portion of its density lying outside of the cloud of the bath [Fig. \ref{Fig:Density_1p4_0p6} (d)]. The density of the latter resembles a distorted TF profile, in sharp contrast to the MF case [Fig. \ref{Fig:Density_1p4_0p6} (a)], where $\rho^{(1),A}$ develops a three-dip structure which suggests a significantly more excited background than the MB case. The distorted TF profile then provides an effective potential for the impurities, which resembles a harmonic trap, as can be seen in Fig. \ref{Fig:Effective_pot2} (c) at later times $t=170$. The absence of any potential wells is a signature of the miscible character of the impurity-medium interactions. In that case a superposition of many excited states is needed in order to properly account for the density profiles similar to those displayed in Fig. \ref{Fig:Density_1p4_0p6} (d) \cite{Theel,Mista3}.

  \subsection{Dynamical response for $\Omega>\omega$}
  
  Turning to larger modulation frequencies, the dynamical response of both components is substantially different from the previous case where $\Omega<\omega$. More precisely, the spontaneously generated patterns emerging in the course of the MF evolution, clearly resemble DB solitons, as is presented in Fig. \ref{Fig:Density_1p4_0p6} (e) and (f) for $\Omega=0.5$. Recall that similar structures have been shown to be nucleated in the reverse driving scenario for strong driving frequencies, however they were shown to be not as robust as here and to form a bound pair [Sec. \ref{Sec:Immiscible}]. Moreover, the oscillation frequency of these structures is much larger than the one associated to the entities in the reverse driving scenario [see Fig. \ref{Fig:Dens_0p2_1p2_2} (a), (b)], and importantly it crucially depends on $\Omega$, as can be easily deduced by inspecting Figs. \ref{Fig:Density_1p4_0p6} (e), (f) [$\Omega=0.5$] and Fig. \ref{Fig:Density_1p4_0p6_2} (a), (b) [$\Omega=1$]. Also, their oscillation amplitude changes with respect to $\Omega$ and in particular it increases from $\Omega=0.5$ to $\Omega=1$ by approximately $59 \%$. Another difference that occurs with the respective structure formation within the MF approach for $\Omega>\omega$ compared to the reverse modulation discussed in Sec. \ref{Sec:Immiscible} [Fig. \ref{Fig:Dens_0p2_1p2_2} (a), (b)], is the existence of a larger amount of excitations, which consequently alter the shape of the pronounced oscillating humps during the time-evolution [Fig. \ref{Fig:Density_1p4_0p6_2} (b)]. For sufficiently long evolution times ($t>300$), these oscillating density humps increase in amplitude and gradually fade away, as a result of the prominent interference processes caused by the miscible nature of the bosonic mixture, as can be seen in Fig. \ref{Fig:Density_1p4_0p6_2} (b). 
  
\begin{figure}[t!]
  	\centering
  	\hspace*{-0.4cm}
  	\includegraphics[width=0.52 \textwidth,keepaspectratio]{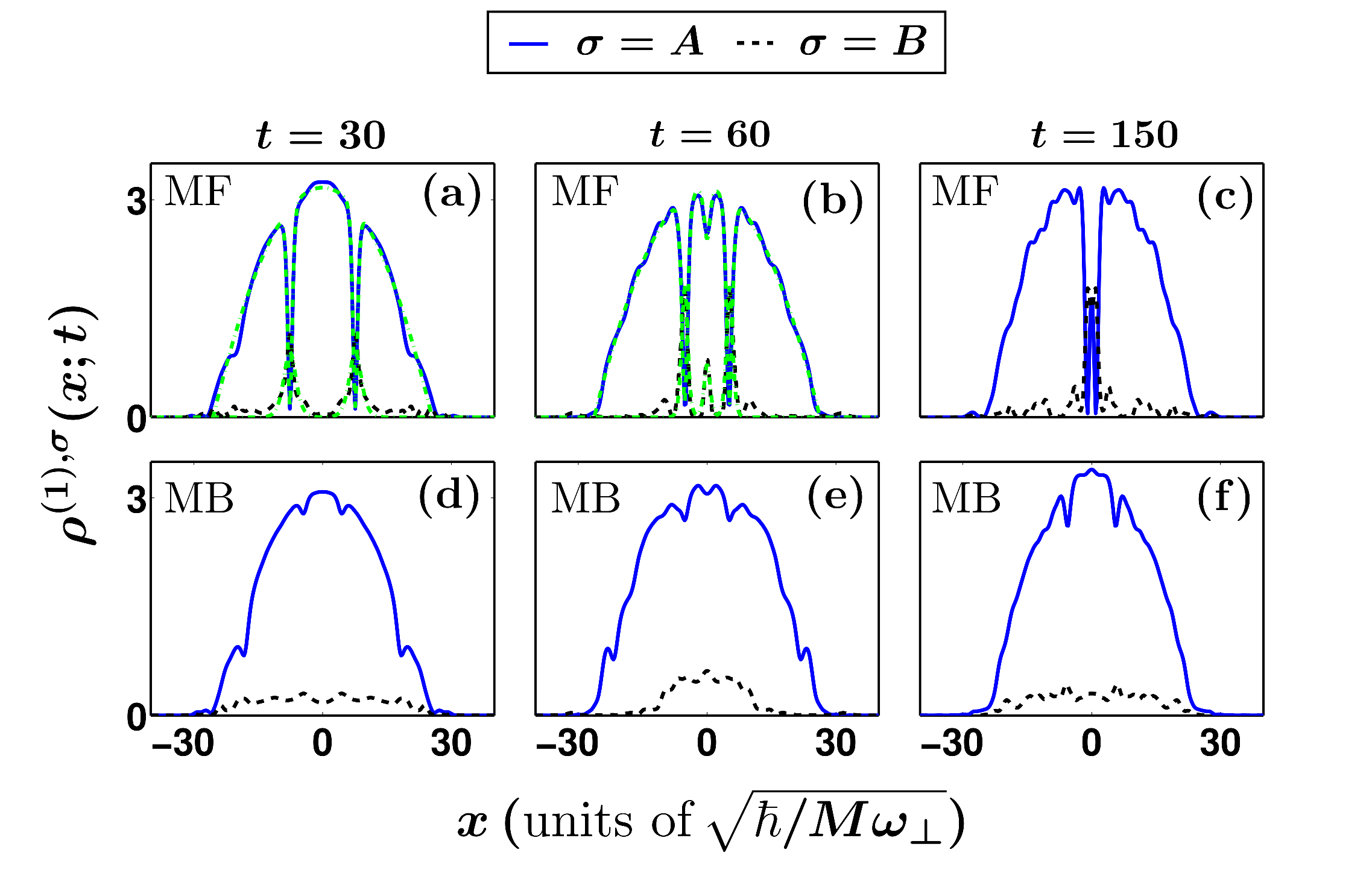}
  	\caption{Profile snapshots of the one-body density of the bath (A) and the impurities (B) following a  modulation of the impurity-medium coupling with $\Omega=0.5$ within the (a)-(c) MF and (d)-(f) MB frameworks. The dashed green lines in (a), (b) present DB soliton fits on the density of both components.}
  	\label{Fig:Snapshots_2}
  \end{figure}

  To further support our argument regarding the character of these structures, we employ the known DB soliton waveforms [Eqs. \eqref{solitons1}, \eqref{solitons2}], denoted by dashed green lines in Figs. \ref{Fig:Snapshots_2} (a), (b). As already mentioned, however, there are excitations on top of $\rho^{(1),B}(x;t)$, which render the fitting of the bright soliton waveform not so accurate. Regarding the bath, the spatiotemporal evolution of its phase [inset of Fig. \ref{Fig:Density_1p4_0p6_2} (a)] displays phase jumps at the positions of the density dips. These jumps being multiples or less than $\pi$, are of course indicative of the presence of moving dark (i.e. gray) solitons \cite{Yan,Lia2,PanosB}. Moreover, the oscillation period of the DB structures that we obtain for $\Omega=1$ is $T^{osc}=112.4$, whereas the theoretical prediction yields $T^{DB}=108.7154$ \cite{Busch,Yan}. This discrepancy is predominantly attributed to the interactions among the solitons and the background excitations of the impurities \cite{PanosB}. Let us finally mention that for a larger pulse duration the period and amplitude of the above-described DB solitons remain almost un-affected while the background becomes more excited because a larger amount of energy is introduced into the system.

  Incorporating correlations, the behavior of the density of the bath and the impurities for $\Omega=0.5$ and $\Omega=1$ [Figs. \ref{Fig:Density_1p4_0p6} (g), (h) and Fig. \ref{Fig:Density_1p4_0p6_2} (c), (d)], is evidently altered from the respective MF time-evolution for $t>5$. Focusing on the impurities, $\rho^{(1),B}(x;t)$ displays initially a density hump for both modulation frequencies close to the trap center, which reflects the immiscible character of the system since at $t=0$ $g^{\textrm{in}}_{AB}=1.4$, and later on it diffuses within the medium suffering enhanced interference phenomena due to the miscible character of the system. The initial density hump subsequently splits, a process which is more prominent in the case of the initial density dip of the bath, as we shall discuss later on [Fig. \ref{Fig:Density_1p4_0p6} (g) and Fig. \ref{Fig:Density_1p4_0p6_2} (c)]. The impurities cloud undergoes a large amplitude breathing motion with frequency $\omega^{br}\simeq 0.157$ for both $\Omega=0.5$ and $\Omega=1$. This frequency is extracted by calculating the impurities position variance, $\braket{(x^B)^2}$ \cite{Giamarchi,Volosniev}. To explain such a breathing frequency, we resort to the effective potential experienced by the impurities due to the presence of the bath [Eq. \eqref{Effective_potential}]. By inspecting the density snapshots of the bosonic environment in Figs. \ref{Fig:Snapshots_2} (d)-(f), the time-averaged profile $\bar{\rho}^{(1),A}(x)=\frac{1}{T}\int_0^T g_{AB}(t) \rho^{(1),A}(x;t)$ smears out small density fluctuations and resembles a TF profile, $\bar{\rho}^{(1),A}(x)=Q(R^2-x^2)\theta(R^2-x^2)$, with $\theta(x)$ being the heaviside function and $T=200$. Therefore, the small density undulations caused by the impurity motion, present in the instantaneous profiles of $\rho^{(1),A}(x;t)$, are now eliminated. We remark that in the case of $\Omega=0.5$ ($\Omega=1$) $\bar{\rho}^{(1),A}$ saturates for $T>195$ ($T>180$). On top of the time-averaged density of the medium there are small density humps at $x=\pm 5$, which will be discussed later on. The effective potential is a deformed harmonic trap with a renormalized frequency $\omega_{eff}=\sqrt{\omega^2-\frac{2Q}{M}}$ \cite{Ferrier,Hannes,Mista3}. Therefore, the corresponding effective breathing frequency is $\omega_{eff}^{br}=2\omega_{eff}=0.1328$ for $\Omega=1$, with a $1\%$ relative deviation for $\Omega=0.5$. The discrepancy between $\omega^{br}$ and $\omega^{br}_{eff}$ arises due to the presence of correlations \cite{Lia3}, which alter the TF profile, and are imprinted as small density humps on top of $\rho^{(1),A}(x;t)$ at $x=\pm 5$ [Fig. \ref{Fig:Density_1p4_0p6} (g) and Fig. \ref{Fig:Density_1p4_0p6_2} (c)].

  \begin{figure}[t!]
  	\centering
  	\hspace*{-0.7cm}
  	\includegraphics[width=0.52 \textwidth,keepaspectratio]{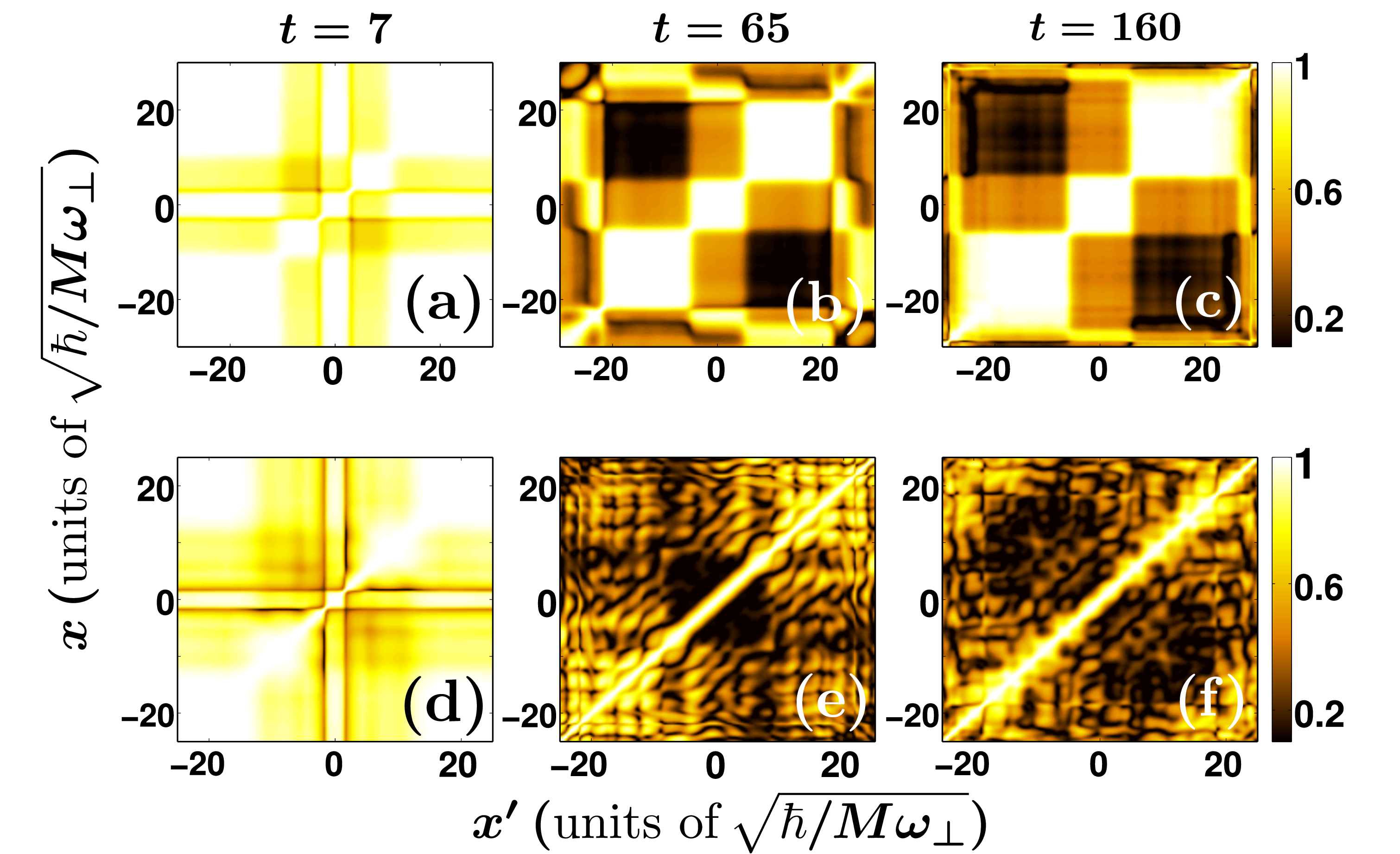}
  	\caption{Snapshots of the first-order coherence of (a)-(c) the bath, $g^{(1),A}(x,x';t)$ and (d)-(f) the impurities, $g^{(1),B}(x,x';t)$. The modulation frequency is $\Omega=0.5$ and all other system parameters are the same as in Fig. \ref{Fig:Density_1p4_0p6} (g), (h).}
  	\label{Fig:Coh_misc}
  \end{figure}
  
  Turning to the dynamical response of the bath $\rho^{(1),A}(x;t)$ features structural changes compared to its MF analogue [Figs. \ref{Fig:Density_1p4_0p6} (e), \ref{Fig:Density_1p4_0p6_2} (a)] especially right after the termination of the impurities-bath interaction modulation, namely at $t \simeq 15$ for $\Omega=0.5$ and at $t \simeq 7$ for $\Omega=1$. Initially ($0<t<2.5$) there is a density dip localized at $x=0$, which splits into two repelling density branches, see the black-dashed ellipses in Fig. \ref{Fig:Density_1p4_0p6} (g) and Fig. \ref{Fig:Density_1p4_0p6_2} (c). Subsequently each of these branches splits further into two shallower density dips, with one travelling towards the edges of the medium and the other one having a significantly smaller amplitude, and remaining almost unaffected throughout time-evolution. The amplitude of these dips increases slightly in time but their position stays the same at $x\simeq \pm 5$, as can be seen in Fig. \ref{Fig:Snapshots_2} (d)-(f). This process together with the splitting of the bright component is reminiscent of the splitting of a quantum DB soliton pair in the presence of correlations, into a fast and a slower moving solitary wave, as reported in Ref. \cite{Lia3}. A similar to the above-described phenomenology occurs for larger modulation frequencies $\Omega$ and therefore also for the case of an impurity-medium interaction quench, see Eq. (\ref{protocol}). 
  However, in the latter case the outer $\rho^{(1),A}$ shallow dips when reaching the trap edges are reflected back and robustly propagate within the medium displaced from the trap center while the inner $\rho^{(1),A}$ dips collide at $t\approx80$ and merge into a single one. Moreover, the impurities exhibit a somewhat larger spatial extent.
  
  The small density dips which remain almost unaffected in the course of the time-evolution manifest themselves in the effective potential [Eq. \eqref{Effective_potential}] for both modulation frequencies [Fig. \ref{Fig:Effective_pot2} (b)]. They form shallow potential wells, and in their positions the impurities showcase small amplitude humps, see for instance Fig. \ref{Fig:Snapshots_2} (f). Apart from these dips, the density of the bosonic medium resembles a TF profile as we have discussed before, and therefore $V_{\textrm{eff}}(x;t)$ is similar to a harmonic trap. Hence a multitude of its eigenstates is needed in order to at least qualitatively account for the impurities density profiles in the course of the evolution [Fig. \ref{Fig:Snapshots_2} (f)].

  \subsection{Correlation patterns and the bunching of impurities}
  
  \begin{figure}[t!]
  	\centering
  	\hspace*{-0.4cm}
  	\includegraphics[width=0.55 \textwidth,keepaspectratio]{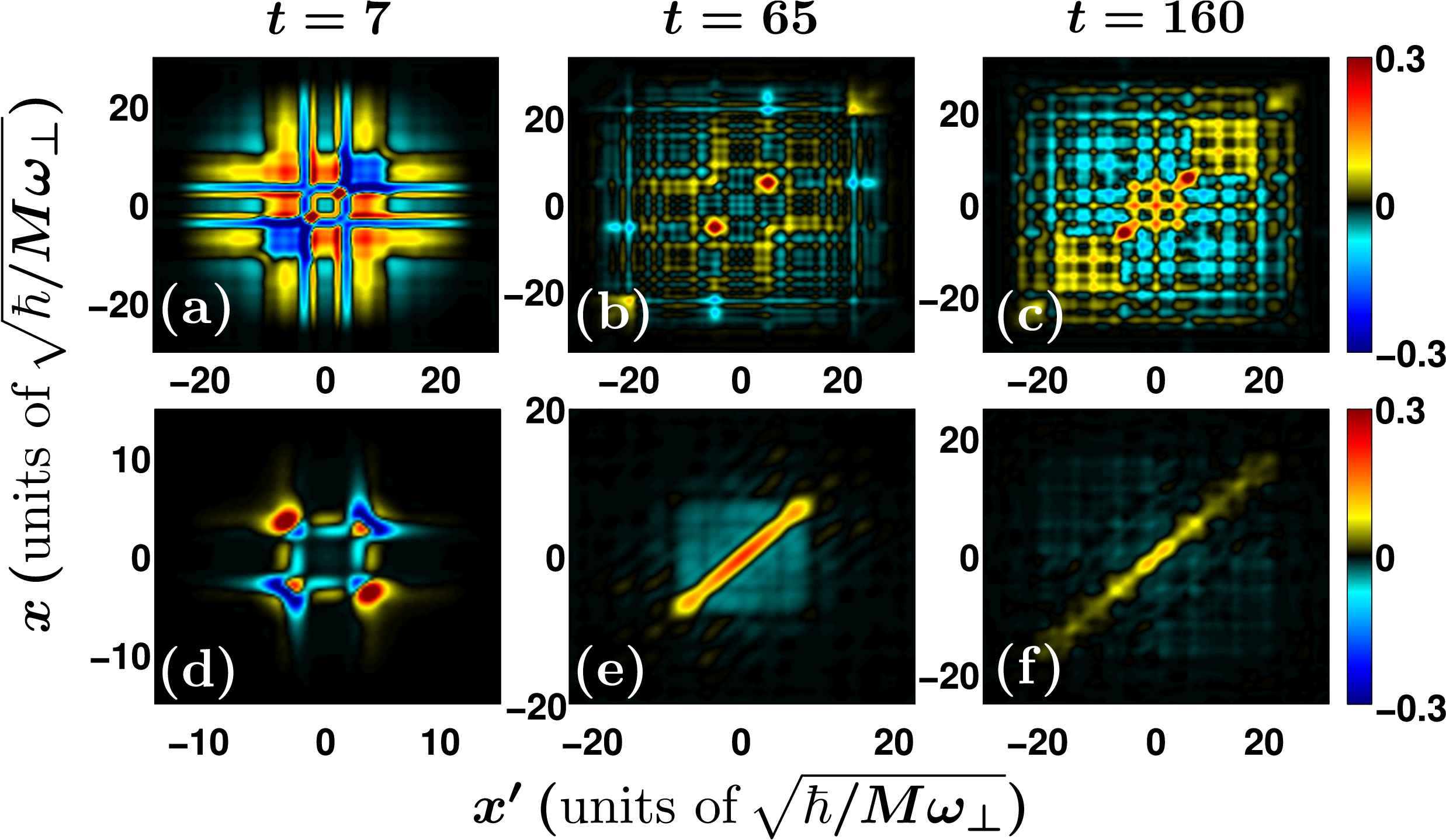}
  	\caption{Instantaneous profiles of the second-order coherence of (a)-(c) the bath particles and (d)-(f) the impurities. In all cases the modulation frequency is $\Omega=0.5$, while all other parameters are the same as in Fig. \ref{Fig:Density_1p4_0p6} (g), (h).}
  	\label{Fig:Coh2_misc}
  \end{figure}

  To expose the role of correlations for the bath and the impurities subsystems, in the driven dynamics to the miscible regime, we employ the first-order coherence function [Eq. \eqref{First_coh}], measuring the underlying coherence losses, and the second-order noise correlation function [Eq. \eqref{Second_coh}], capturing the emergent two-body correlation processes. Initially, the first-order coherence $g^{(1),\sigma}(x,x';t)$ is exemplarily studied during the time-evolution for $\Omega=0.5$ [Fig. \ref{Fig:Coh_misc}]. At the early stages of the dynamics ($0<t<10$) the two narrow density dips at $x\simeq \pm 3$ [see Fig. \ref{Fig:Density_1p4_0p6} (g)] experience a localization trend, see for instance $g^{(1),A}(3.3,-3.6;t=7)\simeq 0.47$ [Fig. \ref{Fig:Coh_misc} (a)]. At later evolution times [Fig. \ref{Fig:Coh_misc} (b), (c)], coherence is almost completely lost for the two symmetric spatial intervals $D^+=(5,22)$ and $D^-=(-22,-5)$ delimited by the mainly stationary density dips, located at $x \simeq \pm 5$ and the outer edges of the medium cloud, with $g^{(1),A}(-15.9,14.6;t=160) \simeq 0.1$. The aforementioned behavior signals the appearance of Mott correlations meaning that the bath particles tend to be localized in either one of those spatial intervals. Turning to the impurities, we observe that at short evolution times, similarly to the bosonic medium, coherence is significantly reduced between the spatial regions corresponding to the density humps [Fig. \ref{Fig:Density_1p4_0p6} (h) at $t\simeq 7$], with $g^{(1),B}(2.54,-3;t=7)\simeq 0.53$ [Fig. \ref{Fig:Coh_misc} (d)]. Later on, the impurities undergo a breathing motion. Upon contraction of the impurity cloud e.g. at $t=65$ [Fig. \ref{Fig:Density_1p4_0p6} (h)], the impurity particles are localized in either of the two spatial intervals $D_+\simeq(0,12)$ and $D_-\simeq (-12,0)$, with $g^{(1),B}(5,-5.5;t=65)\simeq 0.01$ [Fig. \ref{Fig:Coh_misc} (e)]. However, when the impurity cloud expands e.g. at $t=160$, there is still a loss of coherence between the spatial regions away from the trap center, with $g^{(1),B}(-8.7,7.6;t=160)\simeq 0.15$. It is also worth mentioning that for $\Omega<\omega$ the same qualitative picture holds and there is loss of coherence between the outer spatial regions delimited by the small density dips of the bosonic medium cloud.

  Moving to the investigation of two-body correlations, we invoke the second-order noise correlation function $g^{(2)}(x,x';t)$ [Eq. \eqref{Second_coh}], for the same driving frequency, namely $\Omega=0.5$. Initially, e.g. at $t=7$ there is a probability for two particles of the environment to cluster together in the density dips located at $x \simeq \pm 3$, see e.g. $g^{(2),AA}(2.5,-2.8;t=7)\simeq 0.65$ [Fig. \ref{Fig:Coh2_misc} (a)]. Moreover, anticorrelations built for particles occupying the same spatial regions enclosed by the density dips and the edges of the cloud of the bath, e.g. $g^{(2),AA}(-6.56,-6.56;t=7)\simeq -0.17$. Later on, two particles of the bath residing in the two shallow and almost stationary density dips of $\rho^{(1),A}$ located at $x \simeq \pm 5$ exhibit two-body correlations since $g^{(2),AA}(-5.217,-5.217;t=65)\simeq 0.58$ and $g^{(2),AA}(-5.8-5.8;t=160)\simeq 0.64$ [Figs. \ref{Fig:Coh2_misc} (b) and (c) respectively], while opposite spatial regions between the density dips at $x \simeq \pm 5$ and the edges of the medium cloud are anti-correlated [Fig. \ref{Fig:Coh_misc} (c), $g^{(2),AA}(-9,8.7;t=160)\simeq -0.12$]. Turning to the impurity atoms, we observe that initially two-body correlations build up for particles lying on top of the $\rho^{(1),B}$ density humps at $x\simeq \pm 3$, similarly to the case of the bosonic medium, with $g^{(2),BB}(-3.08,-3.08;t=7)\simeq 0.15$ [Fig. \ref{Fig:Coh2_misc} (d)]. Moreover, anticorrelations develop among impurity atoms occupying each of the two distinct density humps, for instance $g^{(2),BB}(-2.54,2.54;t=7)\simeq -0.21$. At later time instants, impurities cluster \cite{Amir,Mista3,Volosniev} and tend to occupy the same position inside the impurity cloud [Fig. \ref{Fig:Coh2_misc} (e) and (f), $g^{(2),BB}(-0.14,-0.14;t=65)\simeq 0.21$]. The second-order noise correlation acquires small negative values (anticorrelations), when the impurities do not reside in the same position, for instance $g^{(2),BB}(1.74,-6.29;t=160)\simeq -0.01$.

  \section{Impact of the impurities interactions and concentration} \label{Sec:Diff_Imp}
  
  Having addressed the impurity-medium pulse dynamics we now demonstrate its dependence on the number of impurities and the impurity-impurity interaction strength. The remaining system parameters are considered to be the same as in the two previous sections \ref{Sec:Immiscible} and \ref{Sec:Miscible}, i.e. $N_A=100$ while $g_{AA}=1.004$. The impurity-bath interaction strength, $g_{AB}$ is driven first from $g^{\textrm{in}}_{AB}=0.2$ to $g^f_{AB}=1.2$, with modulation frequency $\Omega=1.5$, and subsequently from $g^{\textrm{in}}_{AB}=1.4$ to $g^f_{AB}=0.6$ with $\Omega=1$, according to the pulse protocol introduced in Eq. \eqref{protocol}.
  
  \begin{figure}[t!]
      \centering
      \hspace*{-0.2cm}
      \includegraphics[width=0.51 \textwidth,keepaspectratio]{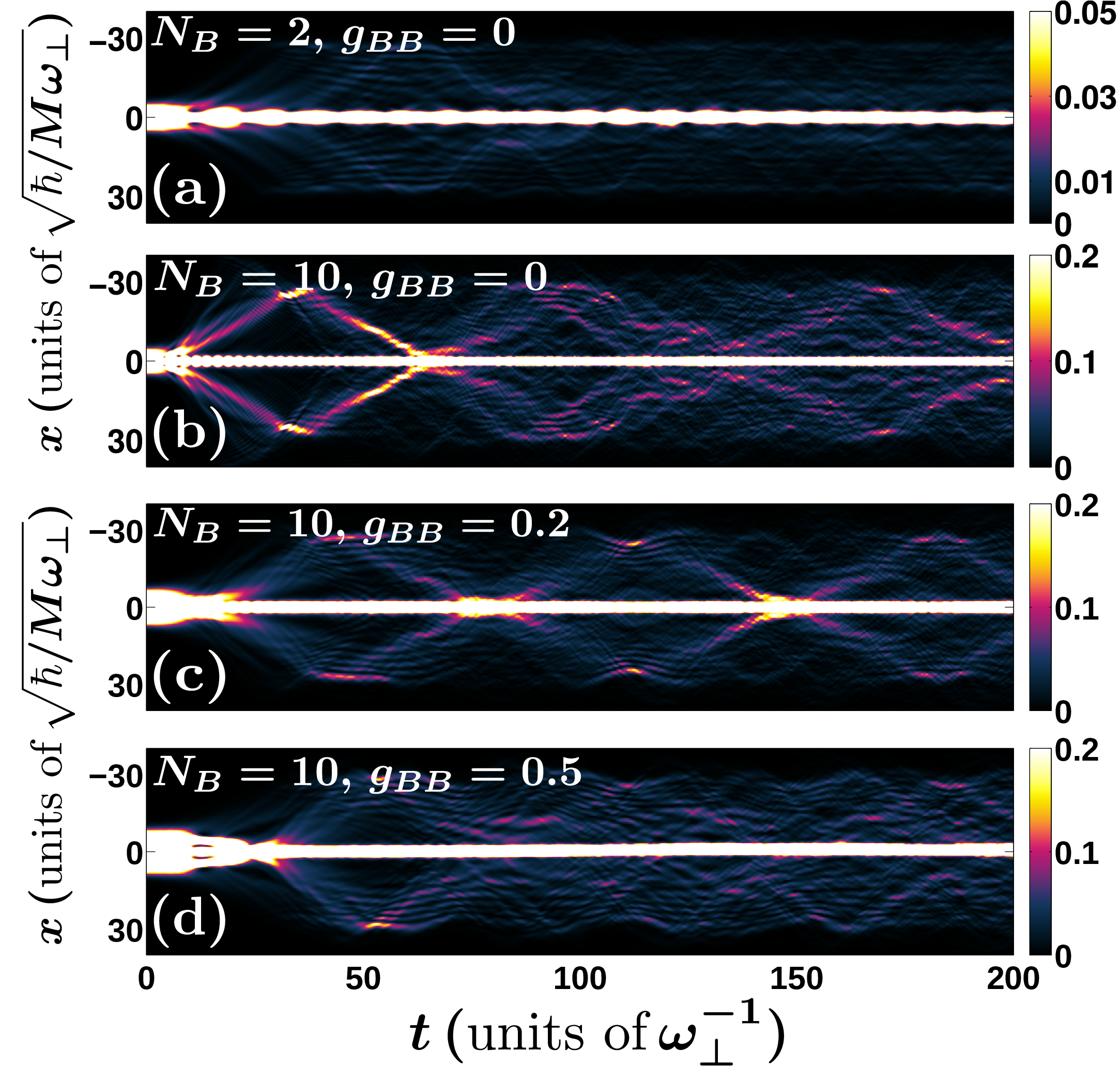}
      \caption{ Time-evolution of the one-body density for (a) $N_B=2$ and (b) $N_B=10$ non-interacting impurity atoms. Density evolution for $N_B=10$ impurities with (c) $g_{BB}=0.2$ and (d) $g_{BB}=0.5$ impurity-impurity interactions. The impurity-bath interaction strength is driven from $g^{\textrm{in}}_{AB}=0.2$ to $g^f_{AB}=1.2$ with frequency $\Omega=1.5$ according to Eq. \eqref{protocol}.}
      \label{Fig:Imp_immiscible}
  \end{figure}
  
  Initially, we explore the impurities dynamical response by considering $N_B=2$ non-interacting ($g_{BB}=0$) ones while driving the impurities-bath interaction strength to the immiscible phase, i.e. from $g^{\textrm{in}}_{AB}=0.2$ to $g^f_{AB}=1.2$ exemplarily with $\Omega=1.5$. The two impurities display mainly a Gaussian profile during the time-evolution and reside around the trap center [Fig. \ref{Fig:Imp_immiscible} (a)], where a density dip is present in $\rho^{(1),A}(x;t)$ of the bosonic medium. Moreover, faint density branches of $\rho^{(1),B}(x;t)$ are emitted and subsequently disperse within the medium \cite{Koushik}. Employing the effective potential picture [Eq. \eqref{Effective_potential}], we deduce that the two particles occupy its ground state with a probability of $93 \%$. For an increasing number of impurities, the time-evolved density of $N_B=10$ non-interacting ones [Fig. \ref{Fig:Imp_immiscible} (b)] is different from the density of $N_B=10$ interacting impurity atoms [Fig. \ref{Fig:Dens_0p2_1p2_2} (d)]. Indeed, for $g_{BB}=0$ there are no prominent outer density humps but rather fragmented faint ones, which after their emission from the central branch oscillate back and forth from the trap center to the edges of the bosonic medium diffusing within the latter. Recall that in the case of $N_B=10$ interacting impurities [Fig. \ref{Fig:Dens_0p2_1p2_2} (d)], the corresponding humps, possessing a significant population, travel away from the trap center and remain at the edges of the environment while oscillating with a small amplitude. This distinct behavior is due to the the presence of repulsive impurity-impurity interactions. The central density hump, which corresponds to the ground state of the effective potential, is present both in the interacting and the non-interacting case. The outer faint humps when $g_{BB}=0$ [Fig. \ref{Fig:Imp_immiscible} (b)] refer to higher-lying excited states of $V_{\textrm{eff}}(x;t)$, localized in its respective outer potential wells, which are shallower compared to the ones of $V_{\textrm{eff}}(x;t)$ in the interacting case due to the different shape of $\rho^{(1),A}(x;t)$ [Fig. \ref{Fig:Effective_pot} (a)].

  We then move on to study the effect of impurity-impurity interactions on the dynamics in the presence of the pulse \cite{Koushik}. The cases of $g_{BB}=0.2$ [Fig. \ref{Fig:Imp_immiscible} (c)] and $g_{BB}=0.5$ [Fig. \ref{Fig:Imp_immiscible} (d)] with $N_B=10$ feature a similar dynamical behavior to the non-interacting ($g_{BB}=0$) scenario [Fig. \ref{Fig:Imp_immiscible} (b)]. Upon increasing $g_{BB}$, the ground state of $\rho^{(1),B}(x;t)$ exhibits a larger spatial extent due to the stronger repulsion, see e.g. Fig. \ref{Fig:Imp_immiscible} (d) with $g_{BB}=0.5$ and Fig. \ref{Fig:Dens_0p2_1p2_2} (d) with $g_{BB}=0.9544$ and the impurities ground state displays a TF profile. The dynamical response of the impurities as quantified by $\rho^{(1),B}(x;t)$ is very similar for $g_{BB}=0$ and $g_{BB}=0.2$ in the sense that there exist faint emitted density branches that oscillate back and forth between the edges of the bosonic bath and $x=0$. These are emitted at later evolution times for a stronger $g_{BB}$, e.g. $g_{BB}=0.2$ [Fig. \ref{Fig:Imp_immiscible} (c)] compared to $g_{BB}=0$ [Fig. \ref{Fig:Imp_immiscible} (b)]. After their creation, they immediately disperse within the cloud of the bath while stronger repulsive interactions lead to larger portions of the impurities occupying the outer density branches [Fig. \ref{Fig:Imp_immiscible} (b) and (c)].
  
 \begin{figure}[t!]
     \centering
     \hspace*{-0.2cm}
     \includegraphics[width=0.51 \textwidth,keepaspectratio]{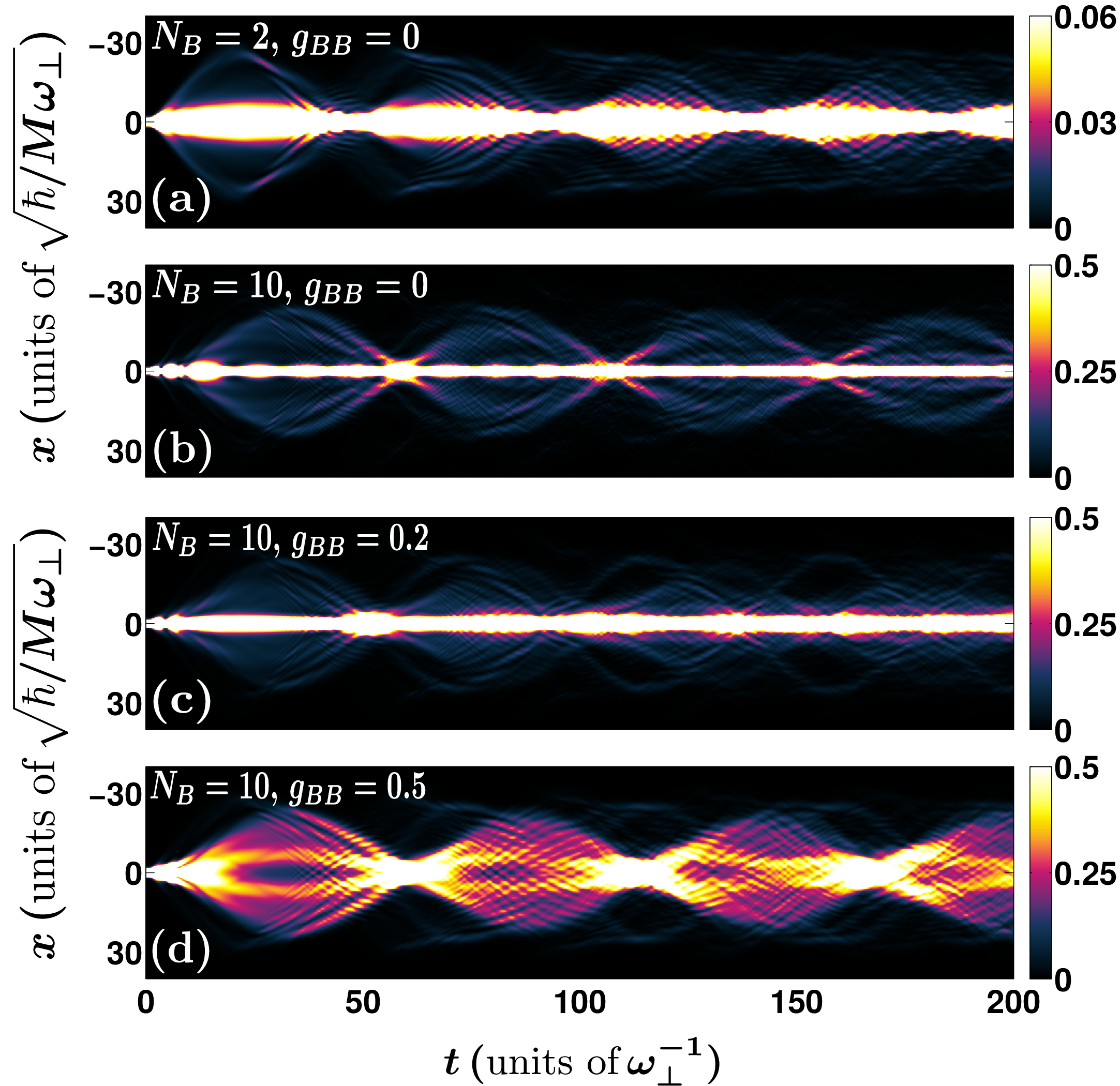}
     \caption{Spatiotemporal evolution of the one-body density for (a) $N_B=2$ and (b) $N_B=10$ non-interacting impurities. Density evolution for $N_B=10$ impurities with (c) $g_{BB}=0.2$ and (d) $g_{BB}=0.5$. The impurity-bath interaction is driven from $g^{\textrm{in}}_{AB}=1.4$ to $g^f_{AB}=0.6$ with the modulation frequency $\Omega=1$ according to Eq. \eqref{protocol}.}
     \label{Fig:Imp_miscible}
 \end{figure}
  
  Furthermore we investigate the impurities response in the reverse pulse scenario, i.e. when the impurity-bath interaction strength drives the system into the miscible phase, with  $g^{\textrm{in}}_{AB}=1.4$ to $g^f_{AB}=0.6$, and $\Omega=1$. Focusing on $N_B=2$ and $N_B=10$ non-interacting impurities [Figs. \ref{Fig:Imp_miscible} (a) and (b) respectively], a breathing motion is apparent with the most prominent frequency being $\omega^{br}=0.157$ in both cases. Note that the latter coincides with $\omega^{br}$ for $N_B=10$ and $g_{BB}=0.9544$ [Fig. \ref{Fig:Density_1p4_0p6_2} (d)]. Moreover, there is a central density hump building upon $\rho^{(1),B}(x;t)$, which is especially pronounced for $N_B=2$ [Fig. \ref{Fig:Imp_miscible} (a)]. This density structure is similar to the case of $N_B=2$ impurities in the reverse pulse scenario, i.e. from the miscible to the immiscible phase [Fig. \ref{Fig:Imp_immiscible} (a)], where the two impurities are predominantly localized around the trap center. In this latter case, however, the two particles exhibit a weaker breathing motion compared to the one triggered by the driven dynamics to the miscible regime [Fig. \ref{Fig:Imp_miscible} (a)]. This is due to the immiscible character of the system following the reverse pulse scenario. Utilizing once more the effective potential picture [Eq. \eqref{Effective_potential}], we can infer that both the $N_B=2$ and $N_B=10$ impurities occupy predominantly its ground state, a result that is manifested by the presence of the central density hump in $\rho^{(1),B}(x;t)$ [Figs. \ref{Fig:Imp_miscible} (a) and (b)].

  As the interactions increase, i.e. $g_{BB}=0.2$ and $g_{BB}=0.5$ [Fig. \ref{Fig:Imp_miscible} (c) and (d) respectively], $\rho^{(1),B}$ shows similar patterns to the one emerging for $g_{BB}=0.9544$, especially for $g_{BB}=0.5$ [Fig. \ref{Fig:Density_1p4_0p6_2} (d)]. Note that this is in contrast to the reverse scenario to the immiscible phase [Fig. \ref{Fig:Imp_immiscible} (c)-(d)], where there is a generic diffusive pattern being apparently different from the localized outer density branches when $g_{BB}=0.9544$ [Fig. \ref{Fig:Density_1p4_0p6_2} (d)]. The central density hump present for $g_{BB}=0$ and $g_{BB}=0.2$ [Figs. \ref{Fig:Imp_miscible} (b) and (c)] corresponds again to the ground state of the respective $V_{\textrm{eff}}(x;t)$, and becomes less prominent for a larger $g_{BB}$ as depicted in Fig. \ref{Fig:Imp_miscible} (d). In the latter case, the effective potential resembles the structure illustrated in Fig. \ref{Fig:Effective_pot2} (b), displaying two shallow wells accounting for the two density humps close to the trap center in the case of $g_{BB}=0.5$ [Fig. \ref{Fig:Imp_miscible} (d)]. Furthermore, both for $g_{BB}=0.2$ and $g_{BB}=0.5$ [Figs. \ref{Fig:Imp_miscible} (c), (d)], the cloud performs a breathing motion, with the most prominent frequency being $\omega^{br}=0.157$, i.e. the same as in the non-interacting case [see Fig. \ref{Fig:Imp_miscible} (b)].

  \section{Summary and Conclusions} \label{Sec:Conclusions}

  We have investigated the non-equilibrium quantum dynamics of few repulsively interacting harmonically trapped bosonic impurities immersed in a MB bosonic bath, subjected to a time-periodic pulse of the impurity-bath interaction strength. Importantly, the effect of the driving frequency on the emergent dynamical response of both components is studied in detail ranging from weak to strong driving. The amplitude of the modulation is large enough to drive the two-component system across its phase separation boundary. In this sense, we examine the driven impurity-medium dynamics from the miscible to the immiscible phase and vice versa.

  Focusing on the driving to the immiscible phase, two distinct response regimes are identified. Namely, if the modulation frequency is smaller than the trapping one, the system transits successively in the course of time from the miscible to the immiscible regime, according to the phase in which it is driven by the impurity-bath coupling. Turning to larger modulation frequencies than the trapping one, DB soliton pairs emerge within the MF approach, which subsequently merge after half of an oscillation period forming a bound state around the trap center. Taking correlations into account, these pairs are expelled towards the edges of the bath cloud, where they equilibrate by performing small-amplitude oscillations. In particular, by comparing the MF and the MB dynamics we conclude that at early evolution times both descriptions yield similar results, but subsequently correlations become important and hence the MF product state does not provide an adequate description. 
  Interestingly, for an increasing modulation frequency we demonstrate that the MF framework is valid only at the very initial stages of the dynamics. The impurity atoms exhibit Mott-like correlations, thus being spatially localized in these outer density branches, which develop two-body correlations among each other. Moreover, a stable density dip (hump) is formed around the trap center in the bath (impurities). This dip splits the bath into two spatial regions which feature two-body correlations. The impurities motion can be intuitively understood in terms of an effective potential picture, unveiling that they predominantly reside in a superposition of its ground and first two excited states.

  In the reverse driving scenario, i.e. following an interaction pulse from the immiscible to the miscible phase we again capture two distinct dynamical regimes, depending on the modulation frequency. For small driving frequencies, the mixture transits consecutively in time from the immiscible to the miscible phase according to the modulation of the impurity-bath coupling. In the time interval that the system lies into its immiscible phase, the impurities reside in a superposition of their lowest-lying effective potential eigenstates. For larger modulation frequencies DB soliton pairs are generated within the MF framework possessing a larger oscillation frequency compared to the previous driving scenario. Incorporating correlations, the impurities perform a breathing motion, whose frequency is in good agreement with the one predicted by their effective potential. We also argue that the dynamical response of the mixture can be well described within the MF approximation only at early evolution times, a result that becomes more pronounced for an increasing modulation frequency where correlation effects are more enhanced. Furthermore, it is found that a multitude of excited eigenstates of their effective potential participate in the dynamics. Regarding the bosonic bath, two small density dips are nucleated, originating from the splitting of the spontaneously generated quantum DB soliton pairs, which are symmetric with respect to the trap center and are almost stable throughout the time-evolution. These dips split the bath into two incoherent parts featuring two-body anticorrelations.

  The role of different impurity particle numbers and impurity-impurity interactions is also explored. It is found that for weak repulsions, the impurities are mainly trapped by the bath around the trap center, occupying predominantly the ground state of their effective potential. This behavior is especially pronounced for two non-interacting particles. By increasing the impurity-impurity interactions or their particle number, weak amplitude emitted density humps form and oscillate between the edges of the cloud of the bath and the trap center. They also exhibit a dispersion within the bath density, mostly for strong repulsions. In particular, when driving the impurity-bath interactions from the immiscible to the miscible phase, it is showcased that the impurities perform a breathing motion with the same prominent frequency regardless of their inherent repulsion.

  The present work can inspire several promising and interesting future research directions. An extension of immediate interest is to consider the two-dimensional analogue of the current setup, where the ejection of correlated jet structures  \cite{Fireworks,Fireworks2,Fireworks3} and the emergence of star-shaped patterns has been reported upon modulating the scattering length \cite{Koushik2}. Additionally, the driving of the impurity-bath coupling strength in the presence of fermionic impurities immersed in a Bose or Fermi gas is an interesting prospect for studying the induced interactions between the impurities and the impact of their flavor in the dynamical response of the system. In a similar vein, the dynamics of bosonic impurities embedded in a fermionic bath with a similar driving protocol, will highlight the role of induced correlations mediated by the fermionic bath \cite{Koushik3}. Certainly, the study of modulated interaction pulses in the presence of dipolar couplings is highly desirable.

\begin{acknowledgments}
G. B. kindly acknowledges financial support by the State Graduate Funding Program Scholarships (HmbNFG).
S. I. M.  gratefully  acknowledges financial support in the framework of the Lenz-Ising Award of the Department of Physics of the University of Hamburg. P.S. is grateful for financial support by the Deutsche Forschungsgemeinschaft (DFG, German Research
Foundation) – SFB-925 – project 170620586.
\end{acknowledgments}

  \appendix
  
   \section{Energy exchange processes} \label{Sec:Appendix_Energy}
   
    \begin{figure}[t!]
  	\centering
  	\hspace*{-0.5cm}
  	\includegraphics[width=0.51 \textwidth,keepaspectratio]{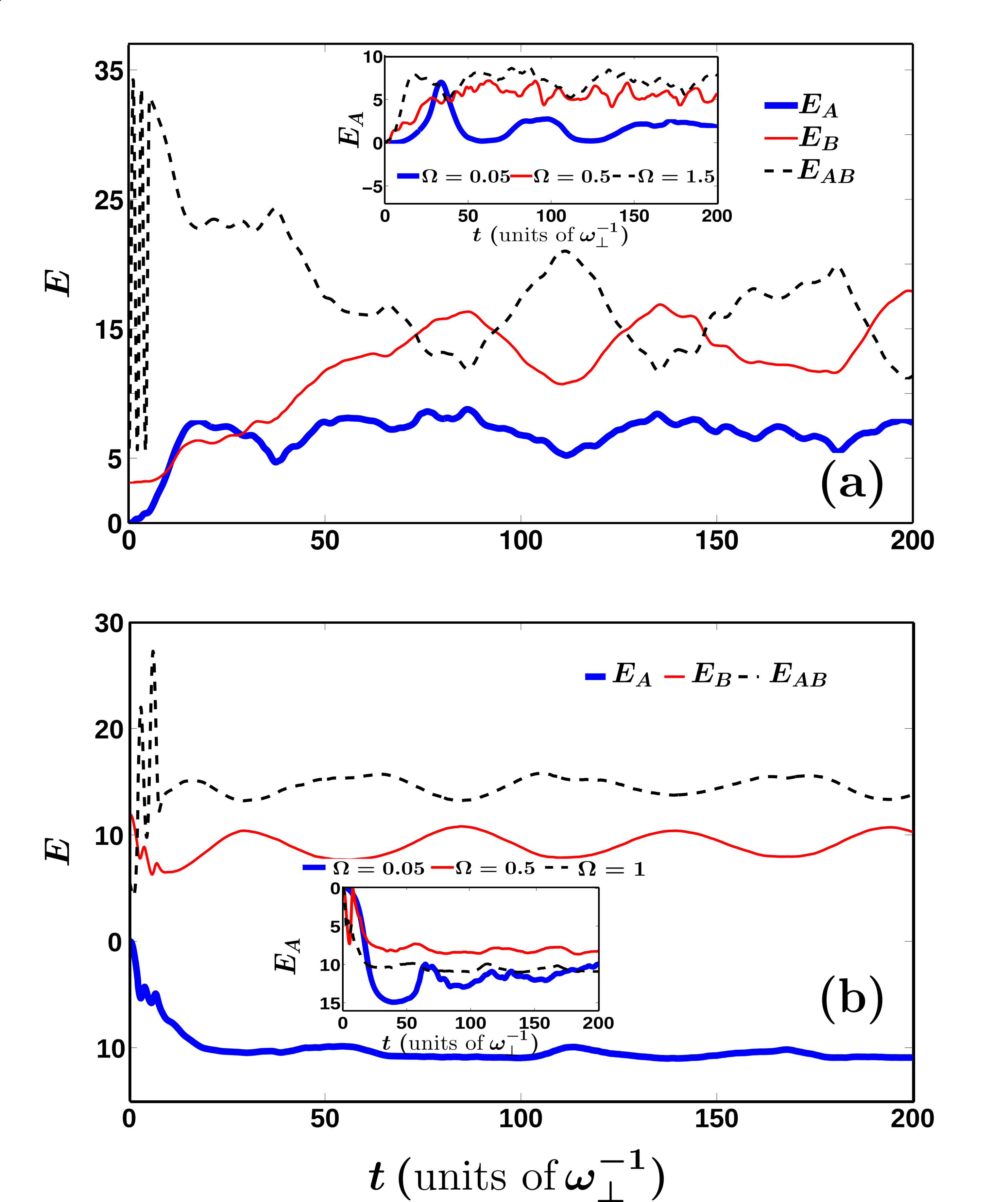}
  	\caption{(a) Energy contributions of the bath ($E_A$), the impurities ($E_B$) and their mutual interaction ($E_{AB}$) following a time-periodic pulse of the impurity-bath coupling from (a) $g^{\textrm{in}}_{AB}=0.2$ to $g^f_{AB}=1.2$ with $\Omega=1.5$. (b) The same as in (a) but for $g^{\textrm{in}}_{AB}=1.4$, $g^f_{AB}=0.6$ and modulation frequency $\Omega=1$. The insets display the energy of the bath for other modulation frequencies (see legend). The energies are given in terms of $\hbar \omega_{\perp}$.}
  	\label{Fig:Energy_1}
  \end{figure}
  
  To elucidate the underlying energy exchange processes between the different components \cite{Giorgini,Diss,Ignuscio} of the system in both driving scenarios addressed in Sections \ref{Sec:Immiscible} and \ref{Sec:Miscible} we invoke the corresponding energy contributions of three different components: namely, the one of the bath ($E_A$), the impurities ($E_B$) and their mutual interactions ($E_{AB}$). In particular the energy of the bath is given by
  
  \begin{eqnarray}
     E_A(t)&=&\braket{\Psi_{\textrm{MB}}(t)| \hat{T}_A+\hat{V}_A(x)+\hat{H}_{AA}|\Psi_{\textrm{MB}}(t)}\\ \nonumber & &-\braket{\Psi_{\textrm{MB}}(0)| \hat{T}_A+\hat{V}_A(x)+\hat{H}_{AA}|\Psi_{\textrm{MB}}(0)}, 
  \end{eqnarray}
  while the energy of the impurities is
  \begin{eqnarray}
     E_B(t)=\braket{\Psi_{\textrm{MB}}(t)| \hat{T}_B+\hat{V}_B(x)+\hat{H}_{BB}|\Psi_{\textrm{MB}}(t)},
  \end{eqnarray}
  and the impurity-medium interaction energy reads
  \begin{eqnarray}
      E_{AB}(t)=\braket{\Psi_{\textrm{MB}}(t) | \hat{H}_{AB}(t)|\Psi_{\textrm{AB}}(t)}.
  \end{eqnarray}
  In these expressions, the kinetic, potential and impurity-bath interaction operators have the form, $\hat{T}_{\sigma}=-\frac{\hbar^2}{2M}\int dx\, \hat{\Psi}^{\sigma \dagger}\frac{d^2}{dx^2}\hat{\Psi}^{\sigma}(x)$, $\hat{V}_{\sigma}(x)=\frac{1}{2}M\omega^2\int dx\, \hat{\Psi}^{\sigma \dagger}(x)x^2 \hat{\Psi}^{\sigma}(x)$, and $\hat{H}_{\sigma \sigma'}(t)=g_{\sigma \sigma'}(t)\int dx\, \hat{\Psi}^{\sigma \dagger}(x) \hat{\Psi}^{\sigma' \dagger}(x)\hat{\Psi}^{\sigma}(x)\hat{\Psi}^{\sigma'}(x)$ respectively with $\sigma=A,B$. Also, $\hat{\Psi}^{\sigma}(x) \, [\hat{\Psi}^{\sigma \dagger}(x)]$ denotes the operator that annihilates [creates] a $\sigma$-species particle at position $x$. Note that the initial energy of the bath which is large due to its substantial spatial extent and particle number, is subtracted in order to render $E_A$ comparable with the other energy contributions.

  Focusing on the driving of the system from the miscible to the immiscible phase with modulation frequency $\Omega=1.5$, the interaction energy $E_{AB}(t)$ initially oscillates according to the quench protocol of Eq. \eqref{protocol} and subsequently decreases [Fig. \ref{Fig:Energy_1} (a)]. Since energy is pumped into the system after the pulse the energy of both components, $E_A$ and $E_B$, increases. The impurities acquire more energy than the bath and this reflects the fact that the outer impurity density branches [Fig. \ref{Fig:Dens_0p2_1p2_2} (d)] reach the edges of the cloud of the bath and remain there while oscillating \cite{Lars,Koushik}. At later time instants $E_B$ features maxima whenever the outer density branches of $\rho^{(1),B}$ [see Fig. \ref{Fig:Dens_0p2_1p2_2} (d) and Fig. \ref{Fig:Energy_1} (a) at $t \approx 85$] tend to the edges of the bath cloud, acquiring thus maximal potential energy, and minima when the $\rho^{(1),B}$ branches approach the trap center [Fig. \ref{Fig:Energy_1} (a), $t \approx 111$]. $E_A$ exhibits a similar behavior and its minima and maxima occur simultaneously with the minima and maxima of $E_B$, since the dips formed in the bath density move in phase with the outer impurity density branches. The impurity-bath interaction energy exhibits out-of-phase oscillations with $E_A$ and $E_B$, which is a manifestation of the energy exchange process between the two components \cite{Lars,Catastrophe}. To infer the behavior of the bath energy with respect to $\Omega$ we present $E_A$ for $\Omega=0.05,0.5$ and $1.5$ [inset of Fig. \ref{Fig:Energy_1} (a)]. Evidently, there is a growth tendency of $E_A$ with increasing $\Omega$, since for larger modulation frequencies more energy is pumped into the system. However for $\Omega=1.5$, $E_A$ is energetically close to the case $\Omega=0.5$. For even larger modulation frequencies, the energy of the bath displays a similar behavior as for $\Omega=1.5$ because for large driving frequencies the effect of the pulse is averaged out \cite{Koushik}. Turning to small $\Omega$ [$\Omega=0.05$ in the inset of Fig. \ref{Fig:Energy_1} (a)], $E_A$ performs small amplitude oscillations, which are in phase with the oscillations of $E_{AB}(t)$ and consequently with the modulation of $g_{AB}(t)$.

  Turning to the reverse pulse scenario, namely from $g^{\textrm{in}}_{AB}=1.4$ to $g^f_{AB}=0.6$ with $\Omega=1$ the impurity-bath interaction energy $E_{AB}(t)$ now increases and afterwards oscillates around a mean value. This behavior is attributed to the fact that the system is driven into the miscible phase where the overlap between the two components is large, compared to the driving to the immiscible regime. The energy of the impurities oscillates around a mean value reflecting their breathing motion [see Fig. \ref{Fig:Density_1p4_0p6_2} (d)], with maxima at the positions where $\rho^{(1),B}(x;t)$ expands  [Fig. \ref{Fig:Density_1p4_0p6_2} (d) at $t \simeq 25$], possessing maximal potential energy, and minima at the locations where $\rho^{(1),B}(x;t)$ contracts [see Fig. \ref{Fig:Density_1p4_0p6_2} (d) at $t \simeq 60$]. Since the impurities energy remains roughly the same and oscillates around a mean value while $E_{AB}$ increases with time due to the miscible character of the system, the bath energy decreases due to energy conservation until $t \simeq 25$ and thereafter oscillates with a small amplitude around a constant value. As can be seen from the inset of Fig. \ref{Fig:Energy_1} (b), $E_A$ becomes negative for other modulation frequencies as well.
  
  \section{Impurities breathing frequency following a pulse to the miscible regime} \label{Sec:Appendix_Breathing}
  
  \begin{figure}[t!]
      \centering
      \includegraphics[width=0.51 \textwidth,keepaspectratio]{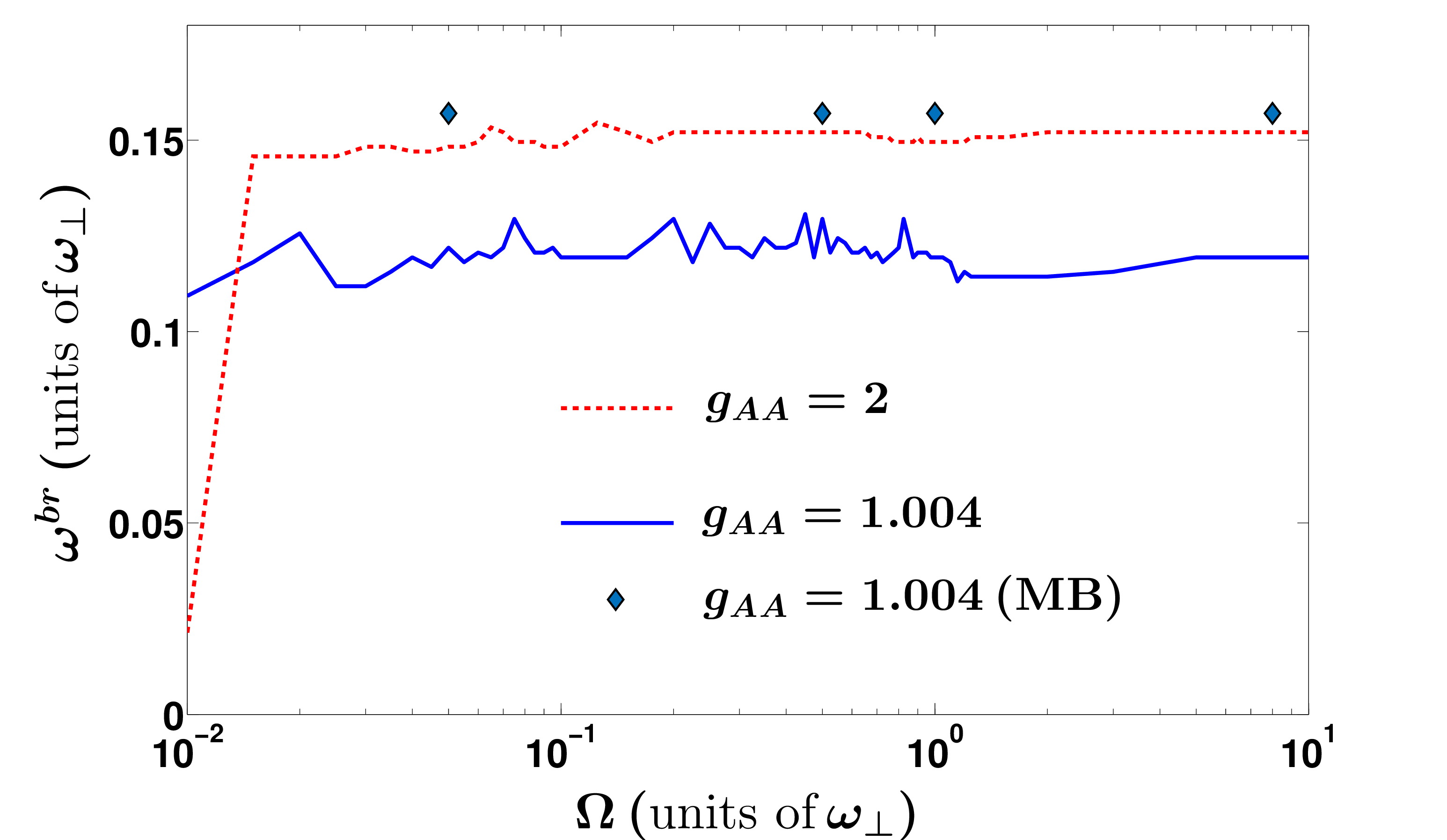}
      \caption{Impurities breathing frequency following a pulse to the miscible phase, i.e. $g^{\textrm{in}}_{AB}=1.4$ and $g^f_{AB}=0.6$ when varying $g_{AA}$ while keeping $g_{BB}=0.9544$ fixed (see legend). The rhombi present the numerically obtained breathing frequency in the MB evolution with $g_{AA}=1.004$.}
      \label{Fig:Breathing_1}
  \end{figure}

  For consistency, let us finally investigate within the MF approximation the role of the driving frequency on the impurities breathing frequency $\omega^{br}$ as the system is driven from the immiscible to the miscible phase, i.e. $g^{\textrm{in}}_{AB}=1.4$ and $g^f_{AB}=0.6$. The breathing frequency is derived by examining the impurities position variance, $\braket{(x_B)^2}$ \cite{Hannes,Giamarchi}. Apart from $\Omega$ the impact of different bath interactions is also explored [Fig. \ref{Fig:Breathing_1}]. By fixing $g_{BB}=0.9544$, $\omega^{br}$ eventually saturates for sufficiently large driving frequencies ($\Omega>5$). More precisely, $\omega^{br}=0.1194$ in the case of $g_{AA}=1.004$ and $\omega^{br}=0.1521$ in the case of $g_{AA}=2$ [Fig. \ref{Fig:Breathing_1}]. Indeed for large $\Omega$, the effect of the pulse is averaged out [Sec. \ref{Sec:Framework} A] and hence the dynamical response of the impurities is unaffected. To explain these breathing frequency values we resort to the effective potential experienced by the impurities for large $\Omega$ [Eq. \eqref{Effective_potential}]. A time-averaging is performed on the medium density, $\bar{\rho}^{(1),A}(x)=\frac{1}{T}\int_0^T g_{AB}(t) \rho^{(1),A}(x;t)$ for sufficiently long evolution times $T$, in order to eliminate small density fluctuations \cite{Diss, Mista3}. For $g_{AA}=1.004$ and $2$, the time-averaged density resembles a TF profile $\bar{\rho}^{(1),A}(x)=Q(R^2-x^2)\theta(R^2-x^2)$, and the breathing frequency is then given by $\omega^{br}=2\sqrt{\omega^2-\frac{2Q}{M}}$ \cite{Hannes}. According to the theoretical predictions \cite{Ferrier,Hannes,Mista3} the latter provides $\omega^{br}=0.13$ and $0.1689$ for $g_{AA}=1.004$ and $g_{AA}=2$ respectively when $\Omega=10$. The relative error of these theoretically anticipated values with the numerically predicted values is of the order of $10 \%$ in both cases. Taking correlations into account for $g_{AA}=1.004$, already from $\Omega=0.05$, $\omega^{br}$ saturates to $0.157$ [rhombi in Fig. \ref{Fig:Breathing_1}], a value well above the MF case where $\omega^{br}=0.1194$, thus suggesting the importance of impurity-impurity correlations \cite{Mista3,Volosniev}.

\end{document}